\documentclass[preprint,amsmath,amssymb,aps,a4]{revtex4-2}
\usepackage{graphicx}
\usepackage{bm}% bold math
\usepackage{txfonts}
\usepackage{hyperref}

\date{\today}
\newcommand{\be}{\begin{eqnarray}}
\newcommand{\ee}{\end{eqnarray}}

\newcommand{\bfb}{{\bf b}_{\perp}}

\newcommand{\bfp}{{\bf p}_{\perp}}

\newcommand{\Dp}{{\bf \Delta}_{\perp}}

\usepackage{xcolor}
\usepackage{rotating}
\begin{document}
%
%\preprint{APS/123-QED}
%
\title{Exploring twist-4 chiral-even GPDs in the light-front quark-diquark model}
\author{Shubham Sharma}
\email{s.sharma.hep@gmail.com}
\affiliation{Department of Physics, Dr. B. R. Ambedkar National Institute of Technology, Jalandhar, 144008, India}
\author{Harleen Dahiya}
\email{dahiyah@nitj.ac.in}
\affiliation{Department of Physics, Dr. B. R. Ambedkar National Institute of Technology, Jalandhar, 144008, India}

\date{\today}% It is always \today, today,
             %  but any date may be explicitly specified
%
\begin{abstract}
In this work, we delve into the realm of quantum chromodynamics (QCD) by calculating twist-4 chiral-even generalized parton distributions (GPDs), within the context of the light-front quark-diquark model (LFQDM), focusing specifically on the intriguing case of zero skewness. To shed light on the behavior of twist-4 chiral-even GPDs, we present comprehensive two-dimensional ($2$-D) and 3-dimensional ($3$-D) plots, mapping their dependence on the longitudinal momentum fraction ($x$) and the momentum transfer ($t$). Furthermore, our research looks into the profound connection between twist-4 chiral-even GPDs and other distribution functions (DFs), including  parton distribution functions (PDFs), transverse momentum-dependent parton distributions (TMDs) and generalized transverse-momentum dependent parton distributions (GTMDs). Additionally, our investigation extends to the exploration of related higher twist form factors (FFs), which play a pivotal role in elucidating the internal structure of hadrons. To provide significant insights into the partons spatial distribution, we have also yielded impact parameter GPD plots.
 
\par
 \vspace{0.1cm}
\end{abstract}
%====================================================
%
\maketitle
%
%\tableofcontents
%
\section{Introduction\label{secintro}}
\begin{figure*}
	\centering
	\begin{minipage}[c]{0.98\textwidth}
		\includegraphics[width=17cm]{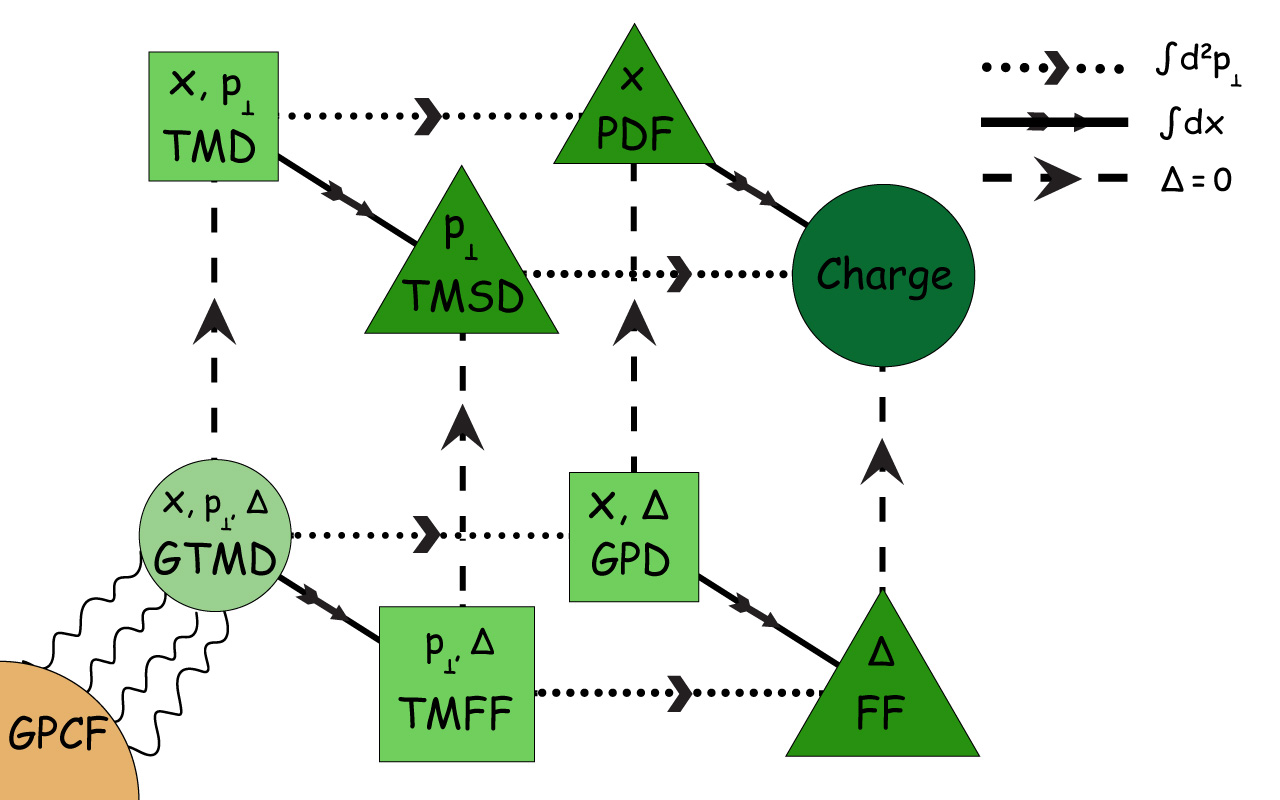}
		\hspace{0.05cm}\\
	\end{minipage}
	\caption{\label{figtree} (Color online) Family tree representation of the generalized Parton Correlation Functions (GPCFs). Different arrows indicate different constraints on the GTMDs. The integration over the quark's transverse momentum $\bfp$ is shown by the dotted line, the solid line signifies the integration over the longitudinal momentum fraction $x$ and the dashed line is used to represent the case of no momentum transfer.
	}
\end{figure*}

Strong nuclear force, one of the four fundamental forces of nature, has been explained by the area of theoretical physics known as quantum chromodynamics (QCD). The QCD theory explains how partons (quarks and gluons), the fundamental particles that make up protons, neutrons, and other hadrons, interact and bind together \cite{Harindranath:1996hq,Brodsky:1997de}. The great deal of emphasis has been imposed by QCD on the study of parton structure because it has broad ramifications for our comprehension of the basic constituents of matter. The longitudinal momentum fraction $x$ represents the fraction of a hadron's longitudinal momentum carried by a particular parton. Understanding $x$ is essential for determining the kinematics of deep inelastic scattering (DIS) \cite{Lai:2010vv}. Accurate knowledge of $x$ is crucial for calculating cross-sections and predicting the outcomes of particle collisions. This leads to the study of mathematical tool known as parton distribution functions (PDFs) \cite{Gluck:1994uf,Collins:1981uw,9803445}, which are linked with parton densities and represent one-dimensional (1-D) structure of the hadron. To obtain more insight on the hadron, two-dimensional (2-D) structures like form factors (FFs) are investigated \cite{Sterman:1997sx}. They provide information about weak or electromagnetic charges and current distribution in a hadron. The three dimensional (3-D) structure of the hadron is studied by transverse momentum-dependent parton distributions (TMDs) \cite{Radici14,Collins81uk,Collins:1981uw,Mulders95,Sivers:1989cc,Kotzinian94,Boer:1997nt} and generalized parton distributions (GPDs) \cite{Mueller:1998fv,Goeke:2001tz,Diehl03,Ji04,Belitsky05,Boffi07,Ji96,Brodsky06,Radyushkin97,Burkardt00,Diehl02,DC05,Ji97,Hagler03,Kanazawa14,Rajan16}. By supplying additional transverse momentum information $\bfp$ to what we already learn from the PDFs, TMDs show a hadron in the momentum space in 3-D \cite{Collins:2003fm,Collins:2007ph,Collins:1999dz,Hautmann:2007uw}. Utilizing the semi-inclusive deep inelastic scattering (SIDIS) and Drell-Yan (DY) processes, this information can be extracted experimentally. Furthermore, information on spin-orbit correlations and the nucleon's angular momentum has been encrypted using TMDs \cite{Bacchetta:2006tn,D'Alesio:2007jt,Burkardt:2008jw,Barone:2010zz,Aidala:2012mv,Cahn:1978se,Konig:1982uk,Chiappetta:1986yg,Collins:1984kg,Sivers:1989cc,Efremov:1992pe,Collins:1992kk,Collins:1993kq,Kotzinian:1994dv,Mulders:1995dh,Boer:1997nt,Boer:1997mf,Boer:1999mm,Bacchetta:1999kz,Brodsky:2002cx,Collins:2002kn,Belitsky:2002sm,Burkardt:2002ks,Pobylitsa:2003ty,Goeke:2005hb,Bacchetta:2006tn,Cherednikov:2007tw,Brodsky:2006hj,Avakian:2007xa,Miller:2007ae,Arnold:2008kf,Brodsky:2010vs,lattice-TMD}. GPDs, which encode information about the parton's longitudinal momentum $x$ and momentum transfer $t$, were experimentally introduced in the context of deeply virtual Compton scattering (DVCS) \cite{Mueller:1998fv,Ji97,Radyushkin:1996nd,Goeke:2001tz,Diehl03,Belitsky05,Boffi07}. The ``mother distributions'' of TMDs and GPDs  are called as generalized transverse momentum-dependent distributions (GTMDs), as it leads to the former under certain kinematic limits \cite{Lorce13}. The theoretical tools that go beyond PDFs, TMDs, GPDs and GTMDs to provide insights into the correlation between partons within hadron are generalized parton correlation functions (GPCFs) \cite{gpcfhalf,Lorce13}.
The integration of GPCFs over the light-cone component of the quark momentum results in GTMDs \cite{Meissner:2008ay,gpcfhalf,Echevarria:2016mrc,Lorce13}. The complete family tree of GPCFs along with other distribution functions like transverse-momentum dependent FFs (TMFFs) and transverse-momentum dependent spin densities (TMSDs) have been shown in Fig. \ref{figtree}. In addition to the distribution functions (DFs) previously mentioned, double parton distribution functions (DPDFs) have gained attention recently because they offer essential details for understanding the hadron's 3-D structure \cite{Kasemets:2017vyh,Rinaldi:2018slz,Diehl:2011yj}. The simultaneous interaction of two partons in high-energy collisions, such as those that take place in multi-parton scattering events at hadron colliders like the Large Hadron Collider (LHC) at CERN, require the application of DPDFs. 
\par

The connection of DVCS process with off-forward parton distributions (OFPDs) was first proposed by Anatoly Radyushkin \cite{Radyushkin:1997yn} and independently by Xiangdong Ji \cite{Ji:1996nm} in late $1990$s. Soon, it was found that OFPDs play a dual role in the sense that they can interpolate between FFs and traditional PDFs  \cite{Burkardt:2000wc}. OFPDs were later popularly known as GPDs and different approaches have been adopted to construct parameterizations and provide successive modeling \cite{Diehl:1998pd,Belitsky:2001ns,Radyushkin:2000uy}. For spin-$\frac{1}{2}$ target, the parameterization of quark GPDs up to twist-4 has been given in Ref. \cite{gpcfhalf}. Model studies of proton GPDs have been done using Nambu-Jona-Lasinio (NJL) model \cite{Freese:2020mcx}, light-front constituent quark model \cite{Lorce:2011dv}, light-cone diquark model \cite{Lu:2010dt}, basis light-front quantization \cite{Kaur:2023lun} and through universal moment parameterization \cite{Guo:2023ahv}. HERMES facility at DESY and experiments at Jefferson Lab have also played a crucial role in advancing our understanding of GPDs and DVCS \cite{Korotkov:2001jk,Xie:2023xkz,Kumericki:2009uq}. From the past few years rigorous method of numerical simulations of QCD on a discrete lattice in Euclidean space-time has been developed by researchers to calculate various observables related to GPDs from first principles \cite{Bhattacharya:2022aob,Ji:2020ect,Constantinou:2020hdm,Lin:2021brq, Bhattacharya:2023nmv}. With high energy, high luminosity future Electron-Ion Colliders (EICs) promises to extract more insight on sea quark and gluon GPDs \cite{Anderle:2021wcy,Aschenauer:2017jsk}.
\par
%================
%PARA 4
%================'
The fundamental purpose of AdS/QCD is to combine two theories, which are Anti-de Sitter space (AdS) in string theory and QCD. As per this theory, a subset of QCD measurements can be transformed into a dual representation within AdS space, where calculations become more manageable. A broad range of fascinating nucleon features, specifically the configuration of the wave function, have been anticipated by the light-front AdS/QCD \cite{Brodsky:2007hb}. It has coherence with the Drell-Yan-West relationship \cite{DY70,West70} and the counting rule for quarks \cite{Maji:2016yqo}. In the light-front quark-diquark model (LFQDM), the proton's composition has been postulated to be made up of an active quark and a diquark observer with a particular mass \cite{Chakrabarti:2019wjx}. From AdS/QCD estimations, the light-front wave functions (LFWFs) have been constructed from the scalar ($S=0$) and axial vector ($S=1$) diquarks with a SU$(4)$ spin-flavor structure \cite{Maji:2016yqo}. We are adequately equipped for the PDFs evolution from its model scale $\mu^2=0.09 ~\mathrm{GeV}^2$ to any required scale upto $\mu^2=10^4~ ~\mathrm{GeV}^2$ in LFQDM.
 \par
The LFWFs within the LFQDM offer a comprehensive view of the likelihood of encountering specific quark and diquark arrangements inside the proton. The LFQDM naturally incorporates the dynamics of confined states, including aspects like confinement and quark-diquark interactions. The LFQDM has effectively portrayed the transverse structure of the proton \cite{Maji:2017bcz}. In the same study, it has been demonstrated that the typical inequalities associated with diquark models are satisfied. Moreover, Ref. \cite{Gurjar:2022rcl} highlights the concordance between the SIDIS spin asymmetries predicted by the LFQDM and the experimental results from HERMES and COMPASS. Quark's orbital angular momentum and the GPD-TMD relations have been calculated, and the results have been contrasted with those of other models that share the same properties \cite{Gurjar:2021dyv}. The LFQDM-derived gravitational FFs (GFFs), $A(Q2)$ and $B(Q2)$, are consistent with the lattice QCD. The D-term FF's qualitative behavior is consistent with information obtained from JLab's DVCS experiments, the lattice QCD and the predictions of numerous other models \cite{Chakrabarti:2020kdc}. The pressure and shear force distributions are in line with other model's predictions \cite{Chakrabarti:2020kdc}. The LFQDM has recently been used to calculate twist-3 \cite{sstwist3} and twist-4 \cite{sstwist4} T-even TMDs. For these higher twist TMDs \cite{sstwist3,sstwist4}, average transverse momenta and average square transverse momenta were calculated, and it was observed that the results closely resembled those of the light-front constituent quark model (LFCQM) and the bag model. The relationships between higher twist TMDs and twist-2 TMDs in this model have been established, and some of them match those in other quark models. The PDF $e(x)$ flavor combination and the most recent CLAS data have also been compared \cite{sstwist3}.
\par
%_____________________________________________________________________________
% tp11 end
%====================================================
To the best of our knowledge, GPDs have not been determined in any model at twist-4. However, research on higher twist TMDs has been done for hadrons \cite{Avakian:2010br,Jakob:1997wg,Lorce:2014hxa,
Pasquini:2018oyz,Kundu:2001pk,Mukherjee:2010iw}. The topic of twist-4 distributions has been discussed in Ref. \cite{Lorce:2014hxa,PhysRevLett.67.552,SIGNAL1997415,PhysRevD.95.074017,ELLIS19821,ELLIS198329,QIU1991105,QIU1991137,PhysRevD.83.054010,liu21,Sharma:2023qgb}. 
In particular, twist-3 \cite{sstwist3} and twist-4 \cite{sstwist4} T-even TMDs have been produced in LFQDM.  Twist-4 GTMDs have also gained a lot of attention recently \cite{Sharma:2023qgb}. In a recent work, the computation of proton's twist-3 GPDs has been carried out using Lattice QCD \cite{Bhattacharya:2023nmv}. Following these factors, it would be interesting to calculate larger twist GPDs in the LFQDM.
\par
The objective of this study is to examine the twist-4 GPDs for the case of the proton using the LFQDM. Specifically, we have decoded the unintegrated quark-quark GPD correlator for the twist-4 Dirac matrix structure and hence obtained the explicit equations for twist-4 GPDs of a proton by comparing it with the parameterization equations. We have derived the explicit equations of GPDs for both the $u$ and $d$ struck quark possibilities from the scalar and vector diquark parts considering the case of skewness $\xi$ being $0$. We show thorough $2$-D and $3$-D graphs, mapping their reliance on the longitudinal momentum fraction $x$ and the momentum transfer $t$, to provide insight into the behavior of twist-4 chiral-even GPDs. To unify its results and relations with other DFs, we explore the related PDFs, TMDs, and GTMDs. This work also includes the analysis of related higher twist FFs, which have been derived from twist-4 GPDs, and are essential for understanding the internal makeup of hadrons. Finally, we have illustrated the results of GPDs in impact parameter space which are obtained via the Fourier transformation of GPDs. 
\par
We have set up our work in the following order: The LFQDM's crucial details, input parameters and other constants are covered in Sec. \ref{secmodel}. The specifics of the twist-4 quark-quark GPD correlator along with the appropriate parameterization equations are described in Sec. \ref{seccor}. In Sec. \ref{secresults}, the explicit equations of the twist-4 GPDs are shown. Relations of twist-4 chiral-even GPDs with GTMDs and TMDs have been explored in Sec. \ref{secrel1} and Sec. \ref{secrel2} sequentially. Analysis of GPDs with the help of 2-D and 3-D plots has been done in Sec. \ref{secdiscussion}. Twist-4 FFs and Fourier-transformed GPD plots are also discussed in this section. The results have finally been concluded in Sec. \ref{seccon}.
%_____________________________________________________________________________
%DONE LAST PARA
%====================================================
%% _____________________________________________________________________________

\section{Light-Front Quark-Diquark Model (LFQDM) \label{secmodel}}
%_____________________________________________________________________________
% tq1 start
%====================================================
The description of a $3$-body system of quarks inside the proton can be considered as a $2$-body system of a quark and a diquark \cite{Maji:2016yqo}. LFQDM  has been proposed based on idea that whenever the proton takes parts in the scattering, it occurs with only one quark and the remaining diquark part will act as a spectator. For the inclusiveness of all the possibilities of struck quark-spectator combination, the SU$(4)$ spin-flavor structure of proton has been assumed to be the blend of isoscalar-scalar diquark singlet $|u~ S^0\rangle$, isoscalar-vector diquark $|u~ A^0\rangle$ and isovector-vector diquark $|d~ A^1\rangle$ states as \cite{Jakob:1997wg,Bacchetta:2008af}
\begin{equation}
	|P; \Lambda^{N} \rangle = C_S|u~ S^0\rangle^{\Lambda^{N}} + C_V|u~ A^0\rangle^{\Lambda^{N}} + C_{VV}|d~ A^1\rangle^{\Lambda^{N}}. \label{PS_state}
\end{equation}
Here, $\Lambda^{N}$ is the nucleon helicity. spin-wise scalar or vector diquark parts have been designated as $S$ or $A=V,~VV$ respectively. Respective isospins of the diquarks have been signified by the $(0)$ or $(1)$ superscripts on them. The coefficients $C_{S}, C_{V}$ and $C_{V V}$ have been derived in Ref. \cite{Maji:2016yqo}, which have been evaluated to be $\sqrt{1.3872}$, $\sqrt{0.6128}$ and $1$ respectively. If $P^+$ is the longitudinal momentum of the proton, the fraction carried by the struck quark of longitudinal momentum $p^+$ is given by $x=p^+/P^+$. Complete light-front momentum co-ordinates of struck quark ($p$) and diquark spectator ($P_X$) can be expressed as
\begin{eqnarray}
	p &&\equiv \bigg(xP^+, p^-,\bfp \bigg)\,,\label{qu} \\
	P_X &&\equiv \bigg((1-x)P^+,P^-_X,-\bfp\bigg). \label{diq}
\end{eqnarray}
The expansion of Fock-state in the two particle case for ${\Lambda^{N}} =\pm 1/2$ for the scalar $|\nu~ S\rangle^{\Lambda^{N}} $ and vector diquark $|\nu~ A \rangle^{\Lambda^{N}}$ in the situation of two particles can be expressed as \cite{majiref25}
\begin{eqnarray}
|\nu~ S\rangle^{\Lambda^{N}} & =& \int \frac{dx~ d^2\bfp}{2(2\pi)^3\sqrt{x(1-x)}} \Bigg[ \psi^{{\Lambda^{N}}(\nu)}_{+}(x,\bfp)\bigg|+\frac{1}{2},~s; xP^+,\bfp\bigg\rangle \nonumber \\
&+& \psi^{{\Lambda^{N}}(\nu)}_{-}(x,\bfp) \bigg|-\frac{1}{2},~s; xP^+,\bfp\bigg\rangle\Bigg],\label{fockSD}\\
|\nu~ A \rangle^{\Lambda^{N}} & =& \int \frac{dx~ d^2\bfp}{2(2\pi)^3\sqrt{x(1-x)}} \Bigg[ \psi^{{\Lambda^{N}}(\nu)}_{++}(x,\bfp)\bigg|+\frac{1}{2},~+1; xP^+,\bfp\bigg\rangle \nonumber\\
&+& \psi^{{\Lambda^{N}}(\nu)}_{-+}(x,\bfp)\bigg|-\frac{1}{2},~+1; xP^+,\bfp\bigg\rangle +\psi^{{\Lambda^{N}}(\nu)}_{+0}(x,\bfp)\bigg|+\frac{1}{2},~0; xP^+,\bfp\bigg\rangle \nonumber \\
&+& \psi^{{\Lambda^{N}}(\nu)}_{-0}(x,\bfp)\bigg|-\frac{1}{2},~0; xP^+,\bfp\bigg\rangle + \psi^{{\Lambda^{N}}(\nu)}_{+-}(x,\bfp)\bigg|+\frac{1}{2},~-1; xP^+,\bfp\bigg\rangle \nonumber\\
&+& \psi^{{\Lambda^{N}}(\nu)}_{--}(x,\bfp)\bigg|-\frac{1}{2},~-1; xP^+,\bfp\bigg\rangle  \Bigg].\label{fockVD}
\end{eqnarray}
Following Eq. (\ref{PS_state}), the flavor index $\nu ~=u$ (for the case of scalar) and   $\nu ~=u,d$ (for the case of vector)). We have  $|\lambda_q,~\lambda_{Sp};  xP^+,\bfp\rangle$ representing the two particle state with quark helicity of $\lambda_q=\pm\frac{1}{2}$ and spectator diquark helicity of $\lambda_{Sp}$. The helicity of spectator for scalar diquark is $\lambda_{Sp}=\lambda_{S}=0$ (singlet) and that for vector diquark is $\lambda_{Sp}=\lambda_{D}=\pm 1,0$ (triplet).
With the possibility of the diquarks being a scalar or a vector, the LFWFs \cite{Maji:2017bcz} have been listed when ${\Lambda^{N}}=\pm1/2$ in Table \ref{tab_LFWF}.
\begin{table}[h]
	\centering % used for centering table
	\begin{tabular}{ |p{1.5cm}|p{1.4cm}|p{1.2cm}|p{1.8cm} p{4.0cm}|p{1.8cm} p{4.0cm}|  }
		%  \hline
		%  \multicolumn{8}{|c|}{Model Parameters corresponding to up \& down quarks } \\
		\hline
		&~~$\lambda_q$~~&~~$\lambda_{Sp}$~~&\multicolumn{2}{c|}{LFWFs for ${\Lambda^{N}}=+1/2$} & \multicolumn{2}{c|}{LFWFs for ${\Lambda^{N}}=-1/2$}\\
		\hline
		~~$\rm{Scalar}$&~~$+1/2$~~&~~$~~0$~~&~~$\psi^{+(\nu)}_{+}(x,\bfp)$~&~~$=~N_S~ \varphi^{(\nu)}_{1}$~~&~~$\psi^{-(\nu)}_{+}(x,\bfp)$~&~~$=~N_S \bigg(\frac{p^1-ip^2}{xM}\bigg)~ \varphi^{(\nu)}_{2}$~~  \\
		&~~$-1/2$~~&~~$~~0$~~&~~$\psi^{+(\nu)}_{-}(x,\bfp)$~&~~$=~-N_S\bigg(\frac{p^1+ip^2}{xM} \bigg)~ \varphi^{(\nu)}_{2}$~~&~~$\psi^{-(\nu)}_{-}(x,\bfp)$~&~~$=~N_S~ \varphi^{(\nu)}_{1}$~~   \\
%		~~S No.~~&~~$\lambda_q$~~&~~$\lambda_D$~~&\multicolumn{2}{c|}{LFWFs for ${\Lambda^{N}}=+1/2$} & \multicolumn{2}{c|}{LFWFs for ${\Lambda^{N}}=-1/2$}\\
		\hline
		&~~$+1/2$~~&~~$+1$~~&~~$\psi^{+(\nu)}_{+~+}(x,\bfp)$~&~~$=~~N^{(\nu)}_1 \sqrt{\frac{2}{3}} \bigg(\frac{p^1-ip^2}{xM}\bigg)~  \varphi^{(\nu)}_{2}$~~&~~$\psi^{-(\nu)}_{+~+}(x,\bfp)$~&~~$=~~0$~~  \\
		~~~~&~~$-1/2$~~&~~$+1$~~&~~$\psi^{+(\nu)}_{-~+}(x,\bfp)$~&~~$=~~N^{(\nu)}_1 \sqrt{\frac{2}{3}}~ \varphi^{(\nu)}_{1}$~~&~~$\psi^{-(\nu)}_{-~+}(x,\bfp)$~&~~$=~~0$~~   \\
		~~$\rm{Vector}$&~~$+1/2$~~&~~$~~0$~~&~~$\psi^{+(\nu)}_{+~0}(x,\bfp)$~&~~$=~~-N^{(\nu)}_0 \sqrt{\frac{1}{3}}~  \varphi^{(\nu)}_{1}$~~&~~$\psi^{-(\nu)}_{+~0}(x,\bfp)$~&~~$=~~N^{(\nu)}_0 \sqrt{\frac{1}{3}} \bigg( \frac{p^1-ip^2}{xM} \bigg)~  \varphi^{(\nu)}_{2}$~~   \\
		&~~$-1/2$~~&~~$~~0$~~&~~$\psi^{+(\nu)}_{-~0}(x,\bfp)$~&~~$=~~N^{(\nu)}_0 \sqrt{\frac{1}{3}} \bigg(\frac{p^1+ip^2}{xM} \bigg)~ \varphi^{(\nu)}_{2}$~~&~~$\psi^{-(\nu)}_{-~0}(x,\bfp)$~&~~$=~~N^{(\nu)}_0\sqrt{\frac{1}{3}}~  \varphi^{(\nu)}_{1}$~~   \\
	&~~$+1/2$~~&~~$-1$~~&~~$\psi^{+(\nu)}_{+~-}(x,\bfp)$~&~~$=~~0$~~&~~$\psi^{-(\nu)}_{+~-}(x,\bfp)$~&~~$=~~- N^{(\nu)}_1 \sqrt{\frac{2}{3}}~  \varphi^{(\nu)}_{1}$~~   \\
		&~~$-1/2$~~&~~$-1$~~&~~$\psi^{+(\nu)}_{-~-}(x,\bfp)$~&~~$=~~0$~~&~~$\psi^{-(\nu)}_{-~-}(x,\bfp)$~&~~$=~~N^{(\nu)}_1 \sqrt{\frac{2}{3}} \bigg(\frac{p^1+ip^2}{xM}\bigg)~  \varphi^{(\nu)}_{2}$~~   \\
		\hline
	\end{tabular}
	\caption{The LFWFs for both diquark cases when ${\Lambda^{N}}=\pm1/2$, for various values of helicities of smacked quark $\lambda_q$ and the spectator diquark $\lambda_{Sp}$. $N_S$, $N^{(\nu)}_0$ and $N^{(\nu)}_1$ are the normalization constants.}
	\label{tab_LFWF} % is used to refer this table in the text
\end{table}
The general form of LFWFs $\varphi^{(\nu)}_{i}=\varphi^{(\nu)}_{i}(x,\bfp)$ in Table \ref{tab_LFWF} has been obtained from the prediction of soft-wall AdS/QCD, and the establishment of parameters $a^\nu_i,~b^\nu_i$ and $\delta^\nu$ have been followed from Ref. \cite{Maji:2017bcz}. We have
\begin{eqnarray}
\varphi_i^{(\nu)}(x,\bfp)=\frac{4\pi}{\kappa}\sqrt{\frac{\log(1/x)}{1-x}}x^{a_i^\nu}(1-x)^{b_i^\nu}\exp\Bigg[-\delta^\nu\frac{\bfp^2}{2\kappa^2}\frac{\log(1/x)}{(1-x)^2}\bigg].
\label{LFWF_phi}
\end{eqnarray}
Under the exchange $x \rightarrow 1-x$, the wave functions $\varphi_i^\nu ~(i=1,2)$ are not identical, and this asymmetry exists even at the AdS/QCD limit $a_i^\nu=b_i^\nu=0$  and $\delta^\nu=1.0$. These LFWFs are the upgraded variant of the soft-wall AdS/QCD prediction \cite{Gutsche:2013zia}. By using the Dirac and Pauli FF data \cite{Cates:2011pz,Diehl:2013xca}, the fitting of parameters $a_i^{\nu}$ and $b_i^{\nu}$ has been performed at the model scale $\mu_0=0.313{\ \rm GeV}$  \cite{Maji:2016yqo}. The value of parameter $\delta^{\nu}$ has been adopted to be unity for LFQDM \cite{Maji:2016yqo} from AdS/QCD \cite{deTeramond:2011aml}. Apart from these, the normalization constants $N_{i}^{2}$ in Table {\ref{tab_LFWF}} are derived from Ref. \cite{Maji:2016yqo}. For completeness, the model parameters for both struck quark flavors have been listed in Table \ref{tab_par}. The AdS/QCD scale parameter $\kappa$ found in Eq. (\ref{LFWF_phi}) has been given the value $0.4~\mathrm{GeV}$ \cite{Chakrabarti:2013dda,Chakrabarti:2013gra}. In accordance with Ref. \cite{Chakrabarti:2019wjx}, we have assumed the proton mass ($M$) and the constituent quark mass ($m$) to be $0.938~\mathrm{GeV}$ and $0.055~\mathrm{GeV}$, sequentially.

\begin{table}[h]
	\centering % used for centering table
	\begin{tabular}{ |c|c|c|c|c|c|c|c| }
		%  \hline
		%  \multicolumn{8}{|c|}{Model Parameters corresponding to up \& down quarks } \\
		\hline
		~~$\nu$~~&~~$a_1^{\nu}$~~&~~$b_1^{\nu}$~~&~~$a_2^{\nu}$~~&~~$b_2^{\nu}$~~&~~$N_{S}$~~&~~$N_0^{\nu}$~~&~~$N_1^{\nu}$~~   \\
		\hline
		~~$u$~~&~~$0.280\pm 0.001$~~&~~$0.1716 \pm 0.0051$~~&~~$0.84 \pm 0.02$~~&~~$0.2284 \pm 0.0035$~~&~~$2.0191$~~&~~$3.2050$~~&~~$0.9895$~~  \\
		~~$d$~~&~~$0.5850 \pm 0.0003$~~&~~$0.7000 \pm 0.0002$~~&~~$0.9434^{+0.0017}_{-0.0013}$~~&~~$0.64^{+0.0082}_{-0.0022}$~~&~~$0$~~&~~$5.9423$~~&~~$1.1616$~~    \\
		\hline
	\end{tabular}
	\caption{Values for the normalization constants $N_{i}^{2}$ and model parameters.}
	\label{tab_par} % is used to refer this table in the text
\end{table}

%%%%%%%%%%%%%%%%%%%%%%%%%%%%%%%%%%%%%%%%%%%%%%%
\section{GPD correlator and parameterization at Twist-4}\label{seccor}
%%%%%%%%%%%%%%%%%%%%%%%%%%%%%%%%%%%%%%%%%%%%%%%
In this section, we have presented a detailed analysis of the GPD correlator along with its parameterization. The quark-quark GPD correlator for proton can be defined following Ref. \cite{gpcfhalf}, is given as
\begin{eqnarray} 
F^{\nu [\Gamma]}_{[\Lambda^{N_i}\Lambda^{N_f}]}(x,\xi,t)=\frac{1}{2}\int \frac{dz^-}{2\pi} e^{\frac{i}{2}p^+ z^-} 
\langle P^{f}; \Lambda^{N_f} |\bar{\psi} (-z/2)\Gamma \mathcal{W}_{[-z/2,z/2]} \psi (z/2) |P^{i};\Lambda^{N_i}\rangle \bigg|_{z^+=z_\perp=0}\,.
\label{corr}
\end{eqnarray} 
Here, the proton's initial and final states are represented by $|P^{i};\Lambda^{N_i}\rangle $ and $|P^{f}; \Lambda^{N_f}\rangle$, where $\Lambda^{N_i}$ and $\Lambda^{N_f}$ denote their helicities respectively. The operator for the quark field is $\psi~(\bar{\psi})$. The set of variables $x,\xi$ and $t$ are what the GPD correlator is dependent on, where $t= \Delta^2=-\Dp^2$ represents the square of the total momentum transfer at zero skewness, i.e. $\xi=- \Delta^+/2P^+=0$ \cite{gpcfhalf}. As a result, we have expressed the GPD correlator $F^{\nu [\Gamma]}_{[\Lambda^{N_i}\Lambda^{N_f}]}(x,\xi,t)$ as $F^{\nu [\Gamma]}_{[\Lambda^{N_i}\Lambda^{N_f}]}(x,\Dp^2)$ or compactly as $F^{\nu [\Gamma]}_{[\Lambda^{N_i}\Lambda^{N_f}]}$ throughout the remaining portion of the manuscript, where $\Gamma$ represents the twist-4 Dirac $\gamma$-matrices, i.e., $\Gamma=\{\gamma^-,\, \gamma^-\gamma^5,\, i\sigma^{j-} \gamma^5\}$. For convenience we have taken the Wilson line, $\mathcal{W}_{[-z/2,z/2]}$, to be $1$ which guarantees that the associated bilocal quark operator possesses SU$(3)$ color gauge invariance. At this instance, we adhere to convention $z^\pm=(z^0 \pm z^3)$, and the kinematics are presented by  
\begin{eqnarray}
P &\equiv& \bigg(P^+,\frac{M^2+\Dp^2/4}{P^+},\textbf{0}_\perp\bigg)\,,\\
\Delta &\equiv& \bigg(0, 0,\Dp \bigg)\,.
%\Delta &\equiv& (0,0,\Dp)\,,
\end{eqnarray}
The average proton momentum is expressed as $P= \frac{1}{2} (P^{f}+P^{i})$ in the symmetric frame, and momentum transfer is denoted as $\Delta=(P^{f}-P^{i})$. The proton's initial and final four momenta are subsequently expressed as
\begin{eqnarray}
P^{i} &\equiv& \bigg(P^+,\frac{M^2+\Dp^2/4}{P^+},-\Dp/2\bigg)\,,\label{Pp}\\
P^{f} &\equiv& \bigg(P^+,\frac{M^2+\Dp^2/4}{P^+},\Dp/2\bigg)\,. \label{Ppp}
\end{eqnarray}
Eqs. (\ref{qu}) and (\ref{diq}) provide the momenta of the striking quark ($p$) and diquark ($P_X$), respectively.
By using Eq.~(\ref{PS_state}) to place expression of Eqs.~(\ref{fockSD}) and (\ref{fockVD}) in Eq.~(\ref{corr}), one may define the GPD correlator for the scalar and vector diquark elements as overlaps of the LFWFs given in Table {\ref{tab_LFWF}} as
\begin{eqnarray} 
F^{\nu [\Gamma](S)}_{[\Lambda^{N_i}\Lambda^{N_f}]}(x,\Dp^2)&=&\int\frac{C_{S}^{2}}{16\pi^3} \sum_{\lambda^{q_i}} \sum_{\lambda^{q_f}} \psi^{\Lambda^{N_f}\dagger}_{\lambda^{q_f}}(x,\bfp+(1-x)\frac{\Dp}{2})\psi^{\Lambda^{N_i}}_{\lambda^{q_i}}(x,\bfp-(1-x)\frac{\Dp}{2}) \nonumber\\  &&\frac{u^{\dagger}_{\lambda^{q_f}}(x P^{+},\bfp+\frac{\Dp}{2})\gamma^{0} \Gamma u_{\lambda^{q_i}}(x P^{+},\bfp-\frac{\Dp}{2})}{2 x P^{+}}{d^2 \bfp}\,, \label{cors} \\
F^{\nu [\Gamma](A)}_{[\Lambda^{N_i}\Lambda^{N_f}]}(x,\Dp^2)&=&\int\frac{C_{A}^{2}}{16\pi^3} \sum_{\lambda^{q_i}} \sum_{\lambda^{q_f}} \sum_{\lambda^{D}} \psi^{\Lambda^{N_f}\dagger}_{\lambda^{q_f} \lambda^D}(x,\bfp+(1-x)\frac{\Dp}{2})\psi^{\Lambda^{N_i}}_{\lambda^{q_i}\lambda^D}(x,\bfp-(1-x)\frac{\Dp}{2}) \nonumber\\  &&\frac{u^{\dagger}_{\lambda^{q_f}}(x P^{+},\bfp+\frac{\Dp}{2})\gamma^{0} \Gamma u_{\lambda^{q_i}}(x P^{+},\bfp-\frac{\Dp}{2})}{2 x P^{+}}{d^2 \bfp}\,, \label{corv} 
\end{eqnarray} 
where, for successive $u$ and $d$ quarks, $C_A=C_V, C_{VV}$. $u^{\dagger}_{\lambda^{q_f}}(x P^{+},\bfp+\frac{\Dp}{2})\gamma^{0} \Gamma u_{\lambda^{q_i}}(x P^{+},\bfp-\frac{\Dp}{2})$ is a spinor product that corresponds to the twist-4 Dirac matrices.
A thorough description of the numerous Dirac spinor combinations has been provided in Ref. \cite{Harindranath:1996hq,Brodsky:1997de}. Here, $\lambda^{q_i}$ and $\lambda^{q_f}$ represent, respectively, the beginning and end states of the quark helicity. Additionally, there is a summation over the diquark helicity $\lambda^{D}$ for the vector diquark. \par
The correlator for $u$ and $d$ quarks in the LFQDM model is represented by the components denoting the scalar and the vector diquark as
\begin{eqnarray} 
F^{u[\Gamma]}_{[\Lambda^{N_i}\Lambda^{N_f}]}(x,\Dp^2) &=&  ~F^{u[\Gamma](S)}_{[\Lambda^{N_i}\Lambda^{N_f}]}(x,\Dp^2) + ~F^{u[\Gamma](V)}_{[\Lambda^{N_i}\Lambda^{N_f}]}(x,\Dp^2)\,,\\
F^{d[\Gamma]}_{[\Lambda^{N_i}\Lambda^{N_f}]}(x,\Dp^2) &=&  ~F^{d[\Gamma](VV)}_{[\Lambda^{N_i}\Lambda^{N_f}]}(x,\Dp^2)\,.
\end{eqnarray} 
Following Ref. \cite{gpcfhalf}, the GPDs corresponding to the twist-4 Dirac matrices can be parameterized as
\begin{eqnarray}
	F_{[\Lambda^{N_i}\Lambda^{N_f}]}^{[\gamma^-]}
	&=& \frac{M^2}{2(P^+)^3} \, \bar{u}(P^{f}, \Lambda^{N_F}) \, \bigg[
	\gamma^+ \, H_3(x,\Dp^2)
	+ \frac{i\sigma^{+\Delta}}{2M} \, E_3(x,\Dp^2)
	\bigg] \, u(P^{i}, \Lambda^{N_i}) \,,
	\label{parg1}\\
	F_{[\Lambda^{N_i}\Lambda^{N_f}]}^{[\gamma^-\gamma_5]}
	&=& \frac{M^2}{2(P^+)^3} \, \bar{u}(P^{f}, \Lambda^{N_F}) \, \bigg[
	\gamma^+\gamma_5 \, \tilde{H}_3(x,\Dp^2)
	+ \frac{\Delta^+ \gamma_5}{2M} \, \tilde{E}_3(x,\Dp^2)
	\bigg] \, u(P^{i}, \Lambda^{N_i}) \,,
	\label{parg2}\\
	F_{[\Lambda^{N_i}\Lambda^{N_f}]}^{[i\sigma^{j-}\gamma_5]}
	&=& - \frac{i\varepsilon_T^{ij} M^2}{2(P^+)^3} \, \bar{u}(P^{f}, \Lambda^{N_F}) \, \bigg[
	i\sigma^{+i} \, H_{3T}(x,\Dp^2)
	+ \frac{\gamma^+ \Delta_T^i  - \Delta^+ \gamma^i}{2M} \, E_{3T}(x,\Dp^2) \nonumber\\*
	& & + \frac{P^+ \Delta_T^i  - \Delta^+ P_T^i}{M^2} \, \tilde{H}_{3T}(x,\Dp^2)
	+ \frac{\gamma^+ P_T^i  - P^+ \gamma^i}{M} \, \tilde{E}_{3T}(x,\Dp^2)
	\bigg] \, u(P^{i}, \Lambda^{N_i}) \,, \quad
	\label{parg3}
\end{eqnarray}
where, the functions of form $X(x,\Dp^2)$ on the right-hand side are GPDs, are a total of $8$ in number. Out of them, $H_3$, $\tilde{H_3}$, $E_3$ and $\tilde{E_3}$ are chiral-even while the remaining are chiral-odd. In this work, we have been focusing only on chiral-even GPDs. Out of them, $\tilde{E_3}$ does not exist at zero skewness. During the computation of above expressions, we have used the relations $\varepsilon^{0123} = 1$, $\varepsilon_T^{ij} = \varepsilon^{-+ij}$, $\sigma^{k l}=i\left[\gamma^{k}, \gamma^{l}\right] / 2$ and $\sigma^{+ \Delta} =\sigma^{+ i}\Delta_i$. $k, l=$\{$-,+,i,j\}$, where the indices $i$ and $j$ are used to denote transverse directions.
\section{Expressions of twist-4 chiral-even GPDs}\label{secresults}
We have replaced Eq. \eqref{fockSD} and Eq. \eqref{fockVD} with the proper polarizations in Eq. \eqref{corr} via Eq. \eqref{PS_state} to obtain the formulae of the twist-4 chiral-even GPDs for both types of diquarks. By selecting one correlation (for $\Gamma= \gamma^-$ and  $\gamma^-\gamma_5$) from Eqs. {\eqref{parg1}} and {\eqref{parg2}}, and properly combining proton polarization, one can get a given GPD.  Conclusively, $H_3$, $\tilde{H_3}$ and $E_3$ can be expressed as
\begin{eqnarray}
	H_{3}^{\nu}(x,\Dp^2) &=& \frac{(P^+)^2}{2 M^2}\Bigg( ~F^{\nu[\gamma^-]}_{[++]}(x,\Dp^2)+F^{\nu[\gamma^-]}_{[--]}(x,\Dp^2)\Bigg)\, \label{g1.1} ,\\ 
	\tilde{H}_{3}^{\nu}(x,\Dp^2) &=& \frac{(P^+)^2}{2 M^2}\Bigg( ~F^{\nu[\gamma^-\gamma_5]}_{[++]}(x,\Dp^2)-F^{\nu[\gamma^-\gamma_5]}_{[--]}(x,\Dp^2)\Bigg)\, \label{g1.2} ,\\  
	E_{3}^{\nu}(x,\Dp^2) &=& \frac{(P^+)^2}{M \Dp^2} 
	\Bigg(
	\Delta_x
	\Big( ~F^{\nu[\gamma^-]}_{[-+]}(x,\Dp^2)-F^{\nu[\gamma^-]}_{[+-]}(x,\Dp^2)\Big) \nonumber \\ 
	&&+ \iota \Delta_y
	\Big( ~F^{\nu[\gamma^-]}_{[-+]}(x,\Dp^2)+F^{\nu[\gamma^-]}_{[+-]}(x,\Dp^2)\Big)
	\Bigg)
	\,. \label{g1.3}
\end{eqnarray}
We define
\begin{eqnarray}
	T_{ij}^{(\nu)}(x,\bfp,\Dp)&=&\varphi_i^{(\nu) \dagger}(x,\bfp+(1-x)\frac{\Dp}{2}) \varphi_j^{(\nu)}(x,\bfp-(1-x)\frac{\Dp}{2})
	\label{Tij1},
\end{eqnarray}
where, $i,j=1,2$. This can be written as 
\begin{eqnarray}
	T_{ij}^{(\nu)}(x,\bfp,\Dp)&=&T_{ji}^{(\nu)}(x,\bfp,\Dp)\label{Tij2},\\
	\varphi_i^{(\nu)\dagger}(x,\bfp+(1-x)\frac{\Dp}{2})&=&\varphi_i^{(\nu)}(x,\bfp+(1-x)\frac{\Dp}{2})\label{Tij3},
\end{eqnarray}
 as a direct result of Eq. \eqref{LFWF_phi} and Eq. \eqref{Tij1}. Both scalar and vector diquark chiral-even GPD expressions for the twist-4 Dirac matrix structure can be expressed as  
\begin{eqnarray} 
	%	\bigg(\frac{p_\perp^2+m^2}{M^2}\bigg)
	%
	% H3 SCALAR
	%
	H_{3}^{\nu(S)} &=& \int \frac{C_{S}^{2} N_s^2}{16 \pi^3} \frac{1}{x^2 M^2}\Bigg[ \bigg(m^2+\bfp^2-\frac{\Dp^2}{4} \bigg) T_{11}^{\nu}+ \bigg(\Big(m^2+\bfp^2-\frac{\Dp^2}{4} \bigg)\bigg(\bfp^2-(1-x)^2\frac{\Dp^2}{4} \bigg)+(1-x)\bigg(\bfp^2 \Dp^2 \nonumber\\
	&&- (\bfp \cdot \Dp)^2 \Big)\bigg) \frac{T_{22}^{\nu}}{x^2 M^2}+ m (1-x)\Dp^2 \frac{T_{12}^{\nu}}{x M}\Bigg]{d^2 \bfp} , \label{ceg1s}\\
	%
	% H3 VECTOR
	%
	H_{3}^{\nu(A)} &=& \int  \frac{C_{A}^{2}}{16 \pi^3}  \bigg(\frac{1}{3} |N_0^\nu|^2+\frac{2}{3}|N_1^\nu|^2 \bigg)\frac{1}{x^2 M^2}\Bigg[ \bigg(m^2+\bfp^2-\frac{\Dp^2}{4} \bigg) T_{11}^{\nu}+ \bigg(\Big(m^2+\bfp^2-\frac{\Dp^2}{4} \bigg)\bigg(\bfp^2-(1-x)^2\frac{\Dp^2}{4} \bigg)\nonumber\\
	&&+(1-x)\bigg(\bfp^2 \Dp^2 - (\bfp \cdot \Dp)^2 \Big)\bigg) \frac{T_{22}^{\nu}}{x^2 M^2}+ m (1-x)\Dp^2 \frac{T_{12}^{\nu}}{x M}\Bigg]{d^2 \bfp} , \label{ceg1v}\\
	%
	% H3 Tilde Scalar
	\tilde{H}_{3}^{\nu(S)} &=& \int- \frac{C_{S}^{2} N_s^2}{16 \pi^3} \frac{1}{x^2 M^2}\Bigg[ \bigg(m^2-\bfp^2+\frac{\Dp^2}{4} \bigg)T_{11}^{\nu}+ \bigg((1-x)\bigg(\bfp^2 \Dp^2 - (\bfp \cdot \Dp)^2  \bigg)-\Big(m^2-\bfp^2+\frac{\Dp^2}{4}\Big)\nonumber\\
	&&\bigg(\bfp^2-(1-x)^2\frac{\Dp^2}{4} \bigg)\bigg) \frac{T_{22}^{\nu}}{x^2 M^2}+ 4m\bfp^2 \frac{T_{12}^{\nu}}{x M}\Bigg]{d^2 \bfp} , \label{ceg2s}
%	\\
	%
		\end{eqnarray}
	\begin{eqnarray}
	%
	%H3 Tilde VECTOR
	%
	\tilde{H}_{3}^{\nu(A)} &=& \int -\frac{C_{A}^{2}}{16\pi^3}  \bigg(\frac{1}{3} |N_0^\nu|^2-\frac{2}{3}|N_1^\nu|^2 \bigg) \frac{1}{x^2 M^2}\Bigg[ \bigg(m^2-\bfp^2+\frac{\Dp^2}{4} \bigg)T_{11}^{\nu}+ \bigg((1-x)\bigg(\bfp^2 \Dp^2 - (\bfp \cdot \Dp)^2  \bigg)\nonumber\\
	&&-\Big(m^2-\bfp^2+\frac{\Dp^2}{4}\Big)\bigg(\bfp^2-(1-x)^2\frac{\Dp^2}{4} \bigg)\bigg) \frac{T_{22}^{\nu}}{x^2 M^2}+ 4m\bfp^2 \frac{T_{12}^{\nu}}{x M}\Bigg]{d^2 \bfp} ,  \label{ceg2v}\\
	%
	%
	%
	% E3 SCALAR
	%
	E_{3}^{\nu(S)} &=& \int \frac{C_{S}^{2} N_s^2}{8 \pi^3} \frac{1}{x^2 M}\Bigg[m \bigg( T_{11}^{\nu}+ \bigg( \Big(\bfp^2-(1-x)^2\frac{\Dp^2}{4}\Big)-2\Big(\frac{(\bfp^2 \Dp^2 - (\bfp \cdot \Dp)^2}{\Dp^2} \Big) \bigg)\frac{T_{22}^{\nu}}{x^2 M^2}\bigg) \nonumber\\
	&&+ \bigg( 2\Big( \frac{(\bfp^2 \Dp^2 - (\bfp \cdot \Dp)^2}{\Dp^2}  \Big)
	-(1-x)\Big(m^2+\bfp^2-\frac{\Dp^2}{4}\Big) \bigg) \frac{T_{12}^{\nu}}{x M}\Bigg]{d^2 \bfp} ,  \label{ceg3s}\\
	%
	% E3 VECTOR
	%
	E_{3}^{\nu(A)} &=& \int  -\frac{C_{A}^{2}}{8 \pi^3}  \bigg(\frac{1}{3} |N_0^\nu|^2 \bigg)\frac{1}{x^2 M}\Bigg[m \bigg( T_{11}^{\nu}+ \bigg( \Big(\bfp^2-(1-x)^2\frac{\Dp^2}{4}\Big)-2\Big(\frac{(\bfp^2 \Dp^2 - (\bfp \cdot \Dp)^2}{\Dp^2} \Big) \bigg)\frac{T_{22}^{\nu}}{x^2 M^2}\bigg) \nonumber\\
	&&+ \bigg( 2\Big( \frac{(\bfp^2 \Dp^2 - (\bfp \cdot \Dp)^2}{\Dp^2}  \Big)
	-(1-x)\Big(m^2+\bfp^2-\frac{\Dp^2}{4}\Big) \bigg) \frac{T_{12}^{\nu}}{x M}\Bigg]{d^2 \bfp}. \label{ceg3v}
\end{eqnarray}
By combining the scalar and vector diquark components, twist-4 chiral-even GPDs for the $u$ and $d$ quarks in the LFQDM model can be written as
\begin{eqnarray} 
	X^{u}(x,\Dp^2) &=&  ~X^{u(S)}(x,\Dp^2) + ~X^{u(V)}(x,\Dp^2)\,,\label{gtmdu} \\
	X^{d}(x,\Dp^2) &=&  ~X^{d(VV)}(x,\Dp^2)\,\label{gtmdd}.
\end{eqnarray} 
We would like to assert that when twist-2 GPDs are calculated from correlator (i.e. Eq. (\ref{corr})) with the use of its parameterization equations, the GPDs $H$ and $\tilde{H}$ turns out to be same, whereas $E$ comes out to be negative in comparison to the approach given in Ref. \cite{Diehl:2001pm,Boffi07}. The latter approach has been followed for twist-2 GPDs calculation in LFQDM \cite{Maji:2017ill}. As twist-4 distributions have the same parameterization structure as twist-2, therefore same pattern has been observed while following these two approaches. This has also been observed that while following various approaches and frames for calculation \cite{Diehl:2001pm,Boffi07,Maji:2017ill,gpcfhalf}, researchers have taken the liberty to dissolve or compensate the multiplying factor of $2$ in $\tilde{H}$ (or $\tilde{H_3}$ at twist-4 ) while discussing GPD relation with GTMDs or TMDs. Also, the practice of making $E$ positive for $u$ quark is being followed by the absorption of the minus sign is also being observed in a few cases.
\par To avoid any unnecessary confusion, we have followed the correlator-parameterization protocol for symmetric frames uniformly throughout the article and provided the relation between various distribution functions using that.
%
%
%%%%%%%%%%%%%%%%%%%%%%%%%%%%%%%%%%%%%%%%%%%%%%%
\section{Relation with twist-4 GTMDs}\label{secrel1}
%%%%%%%%%%%%%%%%%%%%%%%%%%%%%%%%%%%%%%%%%%%%%%%
To derive the relation between GPDs and GTMDs, we have to understand the structure of GTMD correlator and its parameterization equations. The fully unintegrated quark-quark GTMD correlator for zero skewness $	W^{\nu [\Gamma]}_{[\Lambda^{N_i}\Lambda^{N_f}]}(x, \bfp^2,\Dp^2,\bfp \cdot \Dp)$ is linked with the GPD correlator $F^{\nu [\Gamma]}_{[\Lambda^{N_i}\Lambda^{N_f}]}(x,\Dp^2)$
as \cite{gpcfhalf}
\begin{eqnarray} 
F^{\nu [\Gamma]}_{[\Lambda^{N_i}\Lambda^{N_f}]}(x,\Dp^2)=\int {d^2 \bfp} 
	W^{\nu [\Gamma]}_{[\Lambda^{N_i}\Lambda^{N_f}]}(x, \bfp^2,\Dp^2,\bfp \cdot \Dp).
	\label{gtmdcorr}
\end{eqnarray} 
%(for $\Gamma= \gamma^-$ and  $\gamma^-\gamma_5$)
For various values of Dirac matrix structure $\Gamma= \gamma^-$ and  $\gamma^-\gamma_5$, the quark GTMDs can be projected as \cite{gpcfhalf}
\begin{eqnarray}
	W_{[\Lambda^{N_i}\Lambda^{N_f}]}^{[\gamma^-]}
	&=& \frac{M}{2(P^+)^2} \, \bar{u}(P^{f}, \Lambda^{N_F}) \, \bigg[
	F_{3,1}
	+ \frac{i\sigma^{i+} p_T^i}{P^+} \, F_{3,2}
	+ \frac{i\sigma^{i+} \Delta_T^i}{P^+} \, F_{3,3}   + \frac{i\sigma^{ij} p_T^i \Delta_T^j}{M^2} \, F_{3,4}
	\bigg] \, u(P^{i}, \Lambda^{N_i})
	\,, \label{par1} \nonumber\\  \\
	W_{[\Lambda^{N_i}\Lambda^{N_f}]}^{[\gamma^-\gamma_5]}
	&=& \frac{M}{2(P^+)^2} \, \bar{u}(P^{f}, \Lambda^{N_F}) \, \bigg[
	- \frac{i\varepsilon_T^{ij} p_T^i \Delta_T^j}{M^2} \, G_{3,1}
	+ \frac{i\sigma^{i+}\gamma_5 p_T^i}{P^+} \, G_{3,2}
	+ \frac{i\sigma^{i+}\gamma_5 \Delta_T^i}{P^+} \, G_{3,3} \nonumber\\
	&&+ i\sigma^{+-}\gamma_5 \, G_{3,4}
	\bigg] \, u(P^{i}, \Lambda^{N_i})
	\,. \label{par2}
\end{eqnarray}
The mashup of proton polarization given in  Eq. (\ref{g1.1}), Eq. (\ref{g1.2}) and Eq. (\ref{g1.3}), in case of GTMD, yields in the following relations 
\begin{eqnarray}
 \frac{(P^+)^2}{2 M^2}\Bigg( ~W^{\nu[\gamma^-]}_{[++]} +W^{\nu[\gamma^-]}_{[--]} \Bigg)&=&	F_{3,1}^{\nu}\, , \label{r1.1}\\
	\frac{(P^+)^2}{4 M^2}\Bigg( ~W^{\nu[\gamma^-\gamma_5]}_{[++]} -W^{\nu[\gamma^-\gamma_5]}_{[--]} \Bigg) 
	&=& G_{3,4}^{\nu} \, ,  \label{r1.2}\\
	\frac{(P^+)^2}{M \Dp^2} 
	\Bigg(
	\Delta_x
	\Big( ~W^{\nu[\gamma^-]}_{[-+]} -W^{\nu[\gamma^-]}_{[+-]} \Big) 
	%\nonumber \\ 
	+ \iota \Delta_y
	\Big( ~W^{\nu[\gamma^-]}_{[-+]} +W^{\nu[\gamma^-]}_{[+-]} \Big)
	\Bigg) &=& \bigg[
	F_{3,1}^e 
	- 2~\bigg(
	\frac{\bfp \cdot \Dp}{\Dp^2} \, F_{3,2}^e 
	+ F_{3,3}^e 
	\bigg)
	\bigg]
	\,. \label{r1.3}
\end{eqnarray}
Comparing  Eqs. (\ref{g1.1}), (\ref{g1.2}) and (\ref{g1.3}) respectively with Eqs. (\ref{r1.1}), (\ref{r1.2}) and (\ref{r1.3}) via the use of Eq. (\ref{gtmdcorr}), we get
\begin{eqnarray}
	H_3(x,\Dp^2) & = & \int   \, 
	\bigg[F_{3,1}^e(x,\bfp^2,\Dp^2, \bfp \cdot \Dp)\bigg]~{d^2 \bfp} \,, \label{r2.1} \\
	\tilde{H}_3(x,\Dp^2) & = & \int   \, \bigg[
	2~G_{3,4}^e(x,\bfp^2,\Dp^2, \bfp \cdot \Dp) \bigg]~{d^2 \bfp} \,, \label{r2.2} \\
	E_3(x,\Dp^2) & = & \int   \, \bigg[
	F_{3,1}^e(x,\bfp^2,\Dp^2, \bfp \cdot \Dp)
	- 2~\bigg(
	\frac{\bfp \cdot \Dp}{\Dp^2} \, F_{3,2}^e(x,\bfp^2,\Dp^2, \bfp \cdot \Dp)
	\nonumber\\ &&+ F_{3,3}^e(x,\bfp^2,\Dp^2, \bfp \cdot \Dp)
	\bigg)
	\bigg]{d^2 \bfp} \,. \label{r2.3} 
\end{eqnarray}
These equations relate twist-4 GTMDs with twist-4 chiral-even GPDs.
%%%%%%%%%%%%%%%%%%%%%%%%%%%%%%%%%%%%%%%%%%%%%%%
\section{Relation with twist-4 TMDs}\label{secrel2}
%%%%%%%%%%%%%%%%%%%%%%%%%%%%%%%%%%%%%%%%%%%%%%%
When GTMDs are processed through the limit of no momentum transfer i.e. $\Delta=0$, twist-4 T-even TMDs are obtained \cite{gpcfhalf,sstwist4}
\begin{eqnarray}
	  F_{3,1}^{\nu}(x,\bfp^2,0,0) & = & f_3^{\nu}(x,\bfp^2)
	  \,, \label{tmd1} \\
G_{3,4}^{\nu}(x, \bfp^2,0,0) & = & 	g_{3L}^{\nu}(x,\bfp^2) \, \label{tmd2}.
\end{eqnarray}
Substituting these values in Eqs. (\ref{r2.1}), (\ref{r2.2}) and (\ref{r2.3}), we get
\begin{eqnarray}
	H_3(x,0) & = & \int   \, 
	\bigg[F_{3,1}^e(x,\bfp^2,0, 0)\bigg]~{d^2 \bfp}= \int   \, 
	\bigg[f_{3}(x,\bfp^2)\bigg]~{d^2 \bfp} \,, \label{e:gpd_gtmd_25} \\
	\tilde{H}_3(x,0) & = & \int   \, \bigg[
	2~G_{3,4}^e(x,\bfp^2,0, 0) \bigg]~{d^2 \bfp}=\int   \, \bigg[
	2~g_{3L}(x,\bfp^2) \bigg]~{d^2 \bfp} \,, \label{e:gpd_gtmd_27} \\
	E_3(x,0) & = & \int   \, \bigg[
	F_{3,1}^e(x,\bfp^2,0, 0)
	- 2~F_{3,3}^e(x,\bfp^2,0, 0)
\bigg]~{d^2 \bfp} \nonumber\\
&&= \int   \, \bigg[
	f_{3}(x,\bfp^2)
	-2 F_{3,3}^e(x,\bfp^2,0,0)
	\bigg)
	\bigg]~{d^2 \bfp} \,. \label{e:gpd_gtmd_26} 
\end{eqnarray}
Above are the GPD-TMD relations at twist-4, where the GTMD $F_{3,3}$ does not correspond to any known TMD at the respective limit \cite{sstwist4}.
%_____________________________________________________________________________
% tr4 end
%====================================================

%\subsection{Explicit Expressions of TMDs}

\section{Discussion}\label{secdiscussion}
\subsection{GPDs}\label{ssgpds}
To study the functioning of chiral-even twist-4 GPDs, at zero skewness $\xi$, with simultaneous change in variables $x$ and $\Dp^2$, we have plotted their 2-D and 3-D variation. Firstly, we will go through their overall behavior in 3-D plots and then discuss the more subtle dependence on its variables in 2-D plots.
\subsubsection{3-D Plots}\label{sss3d}
\begin{figure*}
	\centering
	\begin{minipage}[c]{0.98\textwidth}
		(a)\includegraphics[width=7.3cm]{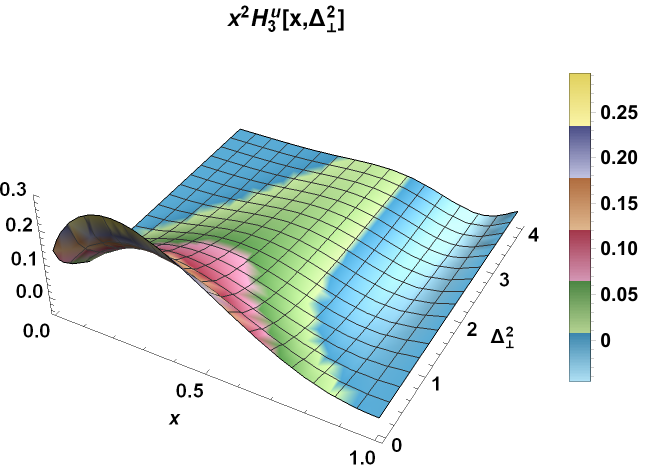}
		\hspace{0.05cm}
		(b)\includegraphics[width=7.3cm]{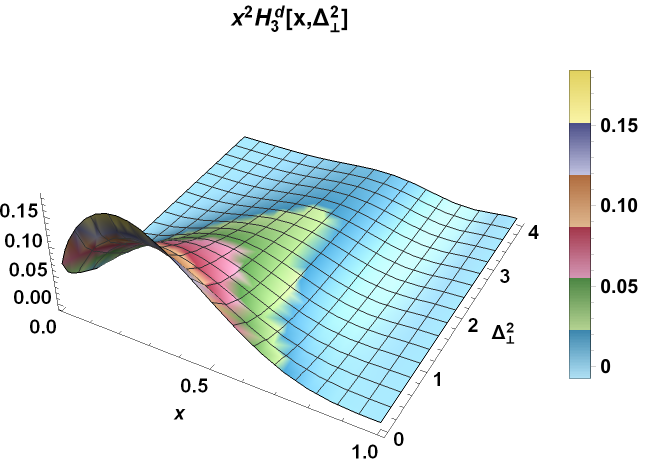}
		\hspace{0.05cm}
		(c)\includegraphics[width=7.3cm]{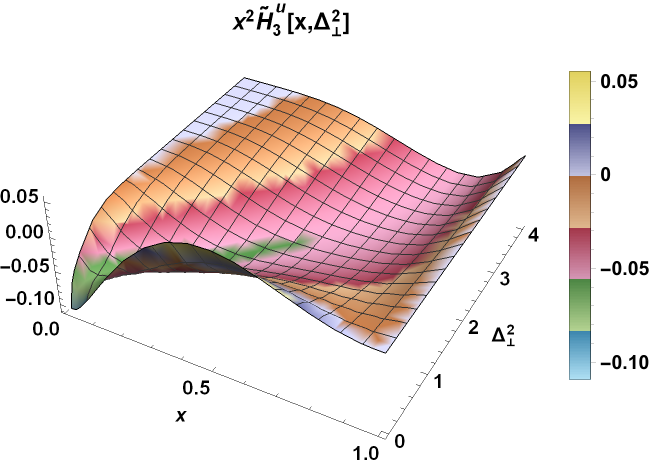}
		\hspace{0.05cm}
		(d)\includegraphics[width=7.3cm]{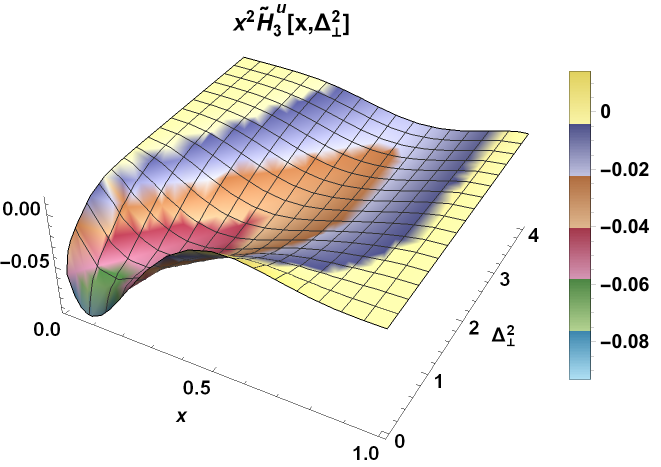}
		\hspace{0.05cm}
		(e)\includegraphics[width=7.3cm]{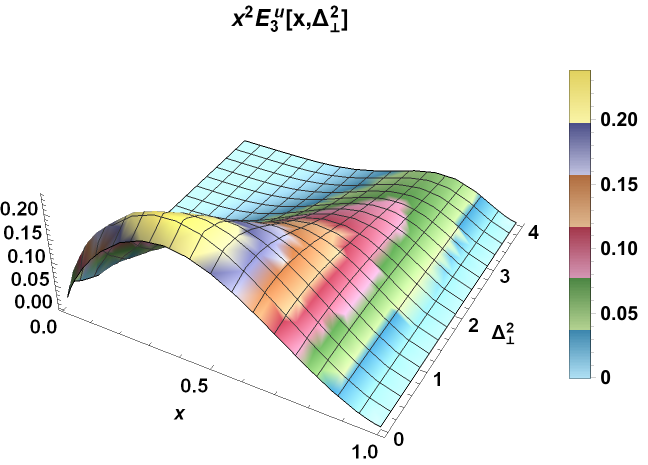}
		\hspace{0.05cm}
		(f)\includegraphics[width=7.3cm]{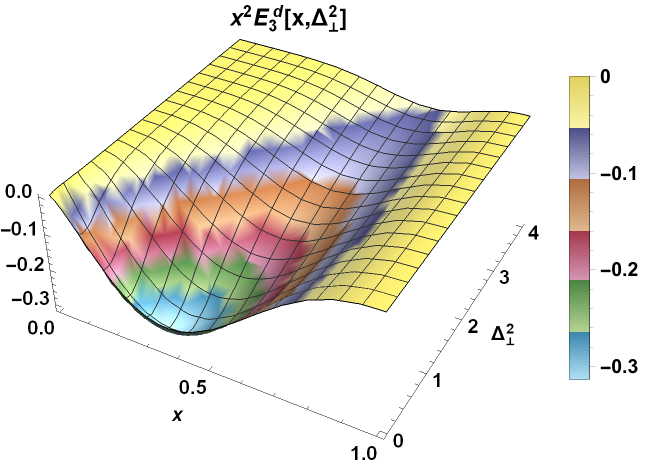}
		\hspace{0.05cm}\\
	\end{minipage}
	\caption{\label{fig3dvxd} (Color online) The chiral-even twist-4 GPDs 		
		$x^2 H_{3}^{\nu}(x,\Dp^2)$,
		$x^2 \tilde{H}_{3}^{\nu}(x,\Dp^2)$ and
		$x^2 E_{3}^{\nu}(x,\Dp^2)$
		plotted with respect to $x$ and $\Dp^2$. The left and right column correspond to $u$ and $d$ quarks sequentially.
	}
\end{figure*}

 In Fig. (\ref{fig3dvxd}), the GPDs $x^2 H_{3}^{\nu}(x,\Dp^2)$, $x^2 \tilde{H}_{3}^{\nu}(x,\Dp^2)$ and $x^2 E_{3}^{\nu}(x,\Dp^2)$ have been plotted with respect to $x$ and $\Dp^2$, where $u$ and $d$ quarks have been shown in the left and right column sequentially. The GPD $x^2 H_{3}^{\nu}(x,\Dp^2)$, plotted in the first row of Fig. (\ref{fig3dvxd}), is associated with the Dirac matrix structure $\gamma^-$ and to the case when proton is longitudinally polarized. The scalar and vector diquark expressions are given by Eq. (\ref{ceg1s}) in Eq. (\ref{ceg1v}) respectively for this GPD. The $T_{11}$ and $T_{22}$ terms which mark the presence of $S$-wave ($L_{z}=0$); and $P$-wave ($L_{z}=\pm1$) contribution is coming from the $T_{12}$ term. The contribution from $S$-wave is more than $80\%$ and it is noteworthy to mention because its twist-2 counterpart, i.e. $H^{\nu}(x,\Dp^2)$, comprises only of $S$-wave \cite{Maji:2017ill}. This leads to a similar kind of variation with variables even though their respective expressions are distinctly different. Also, being related to the unpolarized proton, it is symmetric over the quark flavor. 
\par
The second row in Fig. (\ref{fig3dvxd}), contains the longitudinally polarized proton GPD  $x^2 \tilde{H}_{3}^{\nu}(x,\Dp^2)$, possessing the Dirac matrix structure $\gamma^-\gamma_5$, whose explicit expressions for scalar and vector diquark have been given by Eq. (\ref{ceg2s}) in Eq. (\ref{ceg2v}) sequentially. Here, the $S$-wave contributing terms are $T_{11}$ and $T_{22}$, which act numerically opposite to each other and leading to a significant contribution from $P$-wave i.e., $T_{12}$ term. Even though there is no $P$-wave contribution in the lower twist GPD function $\tilde{H}^{\nu}(x,\Dp^2)$ \cite{Maji:2017ill}, internal conflict in $S$-wave is also prominent there. The symmetry over the quark flavor is due to the proton being longitudinally polarized. 
\par
The final row in Fig. (\ref{fig3dvxd}) contains the transversely polarized proton GPD  $x^2 E_{3}^{\nu}(x,\Dp^2)$, which possesses the Dirac matrix structure $\gamma^-$, whose explicit expressions for both the diquarks have been given by Eq. (\ref{ceg3s}) and Eq. (\ref{ceg3v}). In these expressions, $T_{11}$, $T_{12}$ and $T_{22}$ terms represent the contribution from  
$S$, $P$ and $D$ waves respectively. In this GPD, the relative contribution from $P$-wave is more than $S$-wave which in turn is more than $D$-wave. While studying its twist-2 partner i.e.,  $E^{\nu}(x,\Dp^2)$ \cite{Maji:2017ill}, only $P$-wave contribution has been found. Being transversely polarized, the GPD $x^2 E_{3}^{\nu}(x,\Dp^2)$ is anti-symmetric over the quark flavor. 
Contrary to  $x^2 H_{3}^{\nu}(x,\Dp^2)$ and $x^2 \tilde{H}_{3}^{\nu}(x,\Dp^2)$, the magnitude of $u$ quarks is observed to be less than that for $d$ quarks in GPD $x^2 E_{3}^{\nu}(x,\Dp^2)$. This is because the terms of isoscalar-scalar diquark and isoscalar-vector diquark act opposite to each other in $u$ quarks, whereas in $d$ quarks there is contribution only from isovector-vector diquark.
\subsubsection{2-D Plots}\label{sss2d}
\begin{figure*}
	\centering
	\begin{minipage}[c]{0.98\textwidth}
		(a)\includegraphics[width=7.3cm]{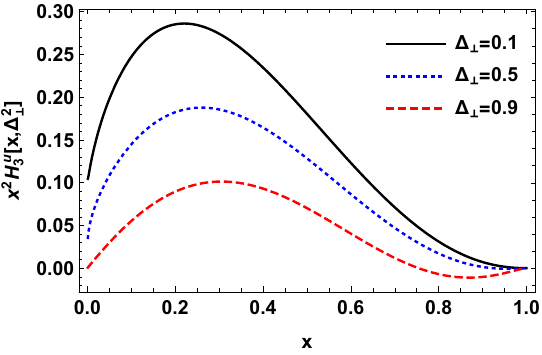}
		\hspace{0.05cm}
		(b)\includegraphics[width=7.3cm]{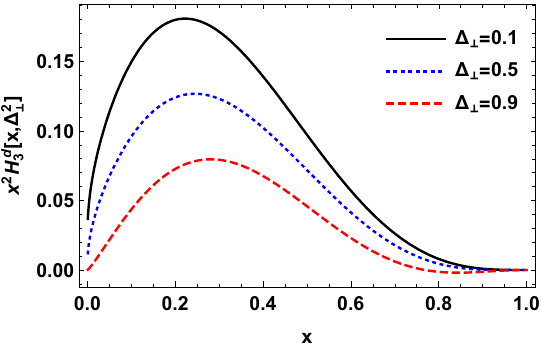}
		\hspace{0.05cm}
		(c)\includegraphics[width=7.3cm]{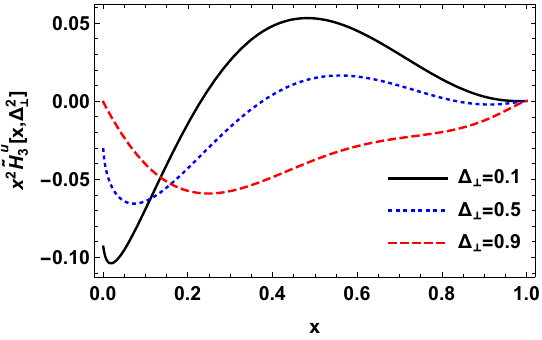}
		\hspace{0.05cm}
		(d)\includegraphics[width=7.3cm]{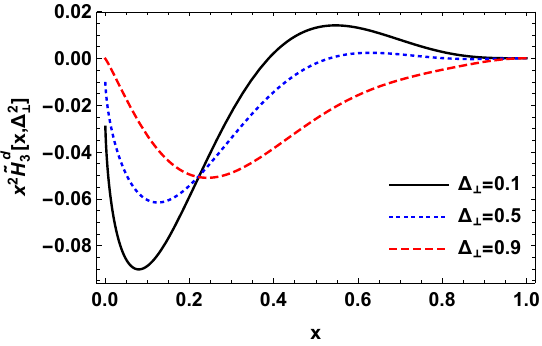}
		\hspace{0.05cm}
		(e)\includegraphics[width=7.3cm]{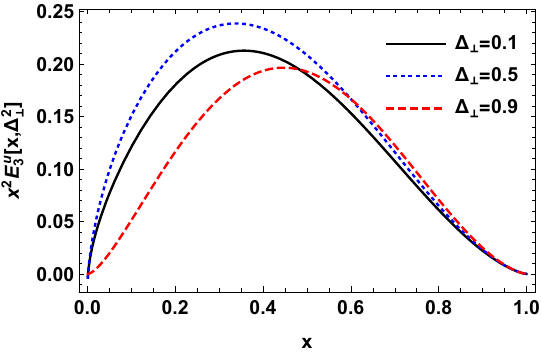}
		\hspace{0.05cm}
		(f)\includegraphics[width=7.3cm]{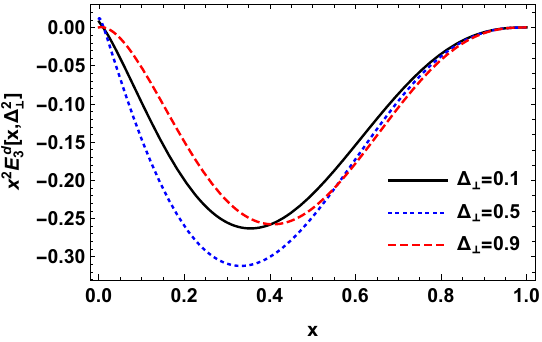}
		\hspace{0.05cm}\\
	\end{minipage}
	\caption{\label{fig2dvx} (Color online)		
	The chiral-even twist-4 GPDs 		
	$x^2 H_{3}^{\nu}(x,\Dp^2)$,
	$x^2 \tilde{H}_{3}^{\nu}(x,\Dp^2)$ and
	$x^2 E_{3}^{\nu}(x,\Dp^2)$
	plotted with respect to $x$ at various fixed values of $\Dp$. The left and right column correspond to $u$ and $d$ quarks sequentially.
	}
\end{figure*}

In Fig. (\ref{fig2dvx}), we have plotted chiral-even twist-4 GPDs $x^2 H_{3}^{\nu}(x,\Dp^2)$, $x^2 \tilde{H}_{3}^{\nu}(x,\Dp^2)$ and $x^2 E_{3}^{\nu}(x,\Dp^2)$ with respect to $x$ at various values of the transverse momentum transfer $\Dp$, for both flavors of the struck quark. 
In the first row of Fig. (\ref{fig2dvx}), the plot of GPD $x^2 H_{3}^{\nu}(x,\Dp^2)$ with respect to $x$ at fix $\Dp$ values has been given and it has been observed that it is very similar to the plot of TMD $f_3^{\nu}(x,\bfp^2)$ \cite{sstwist4}  w.r.t $x$ at fix $\bfp$. This is not surprising as both of these DFs are related to GTMD $F_{3,1}^e(x,\bfp^2,\Dp^2, \bfp \cdot \Dp)$ via Eqs. (\ref{r2.1}) and (\ref{tmd1}). On the same lines, in the second row GPD $x^2 \tilde{H}_{3}^{\nu}(x,\Dp^2)$ has detectable similarity with TMD $g_{3L}(x,\bfp^2)$ \cite{sstwist4} due to its connection GTMD $G_{3,4}^e(x,\bfp^2,\Dp^2, \bfp \cdot \Dp)$, as evident from Eqs. (\ref{r2.2}) and (\ref{tmd2}). In the final row, the GPD $x^2 E_{3}^{\nu}(x,\Dp^2)$ has been observed to be positive (negative) definite over the entire region of the longitudinal momentum fraction $x$ for $u$ ($d$) quarks. 
\par
\begin{figure*}
	\centering
	\begin{minipage}[c]{0.98\textwidth}
		(a)\includegraphics[width=7.3cm]{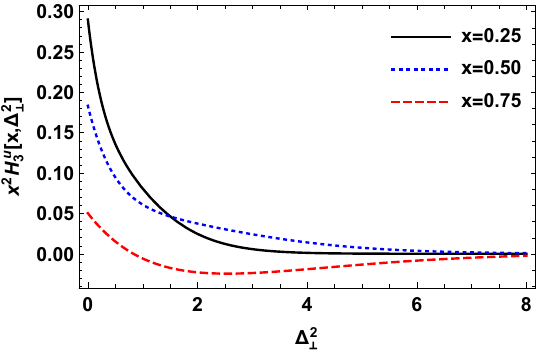}
	\hspace{0.05cm}
	(b)\includegraphics[width=7.3cm]{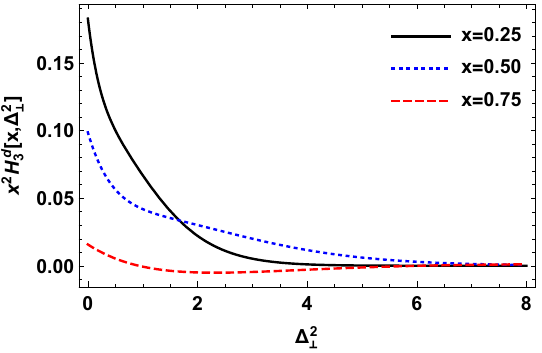}
	\hspace{0.05cm}
	(c)\includegraphics[width=7.3cm]{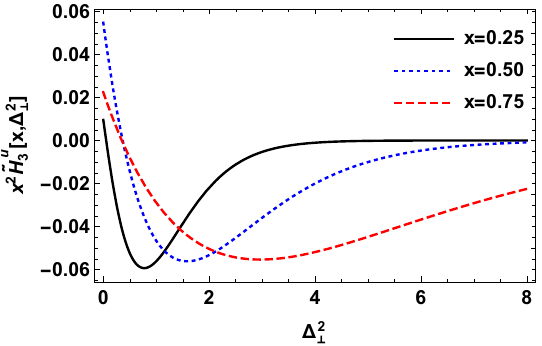}
	\hspace{0.05cm}
	(d)\includegraphics[width=7.3cm]{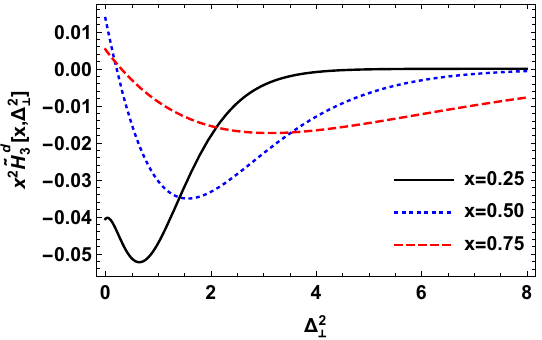}
	\hspace{0.05cm}
	(e)\includegraphics[width=7.3cm]{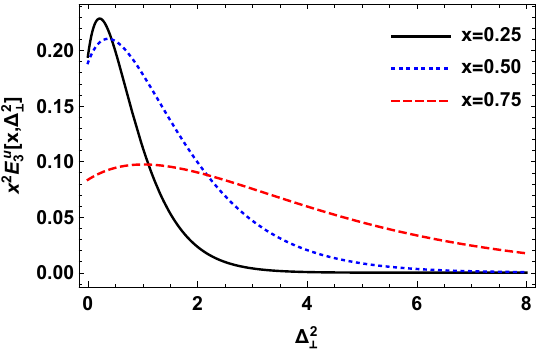}
	\hspace{0.05cm}
	(f)\includegraphics[width=7.3cm]{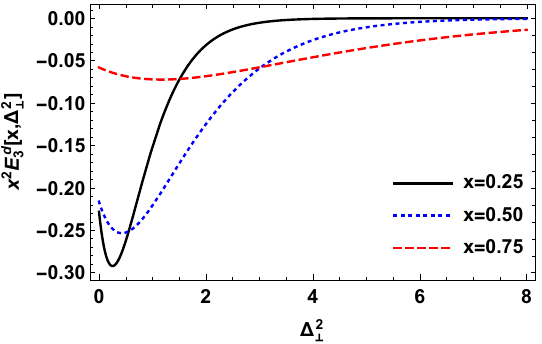}
	\hspace{0.05cm}\\
	\end{minipage}
	\caption{\label{fig2dvd} (Color online) The chiral-even twist-4 GPDs 		
		$x^2 H_{3}^{\nu}(x,\Dp^2)$,
		$x^2 \tilde{H}_{3}^{\nu}(x,\Dp^2)$ and
		$x^2 E_{3}^{\nu}(x,\Dp^2)$
		plotted with respect to $\Dp^2$ at various fixed values of $x$. The left and right column correspond to $u$ and $d$ quarks sequentially.	}
\end{figure*}

In Fig. (\ref{fig2dvd}), the chiral-even twist-4 GPDs 		
$x^2 H_{3}^{\nu}(x,\Dp^2)$, $x^2 \tilde{H}_{3}^{\nu}(x,\Dp^2)$ and $x^2 E_{3}^{\nu}(x,\Dp^2)$ has been plotted with respect to $\Dp^2$ at various fixed values of $x$. The left and right panels correspond to $u$ and $d$ quarks sequentially. This figure is the perfect example of $x \rightarrow 1-x$ asymmetry because in all these plots the curve for longitudinal momentum fraction $x$ at $0.25$ is not the same as that for $0.75$. For fixed longitudinal momentum transfer $x$, as the value of total momentum transfer $\Dp^2$ increases, the amplitude of GPD $x^2 H_{3}^{\nu}(x,\Dp^2)$ decreases for both $u$ and $d$ quarks as shown in Fig. \ref{fig2dvd} (a) and \ref{fig2dvd} (b). Also, the degree of fall in amplitude with $\Dp^2$ is lower for large values of longitudinal momentum fraction $x$. In Fig. \ref{fig2dvd} (c) and \ref{fig2dvd} (d), the GPD $x^2 \tilde{H}_{3}^{\nu}(x,\Dp^2)$ has been plotted with $\Dp^2$, whose peak of minima is observed to be flattened and being drifted to large $\Dp^2$ when the share of its longitudinal momentum fraction is increased. This implies that the dependence on the momentum transfer falls for large $x$. In GPD $x^2 E_{3}^{\nu}(x,\Dp^2)$, the magnitude of extremum decreases and the width on $\Dp^2$ axis increases as the longitudinal momentum fraction carried by struck quark rises as shown in Fig. \ref{fig2dvd} (e) and \ref{fig2dvd} (f). This suggests strong dominance of GPD  $x^2 E_{3}^{\nu}(x,\Dp^2)$ at low $x$ and $\Dp^2$.
\subsection{Form Factors}\label{ssffs}
\begin{figure*}
	\centering
	\begin{minipage}[c]{0.98\textwidth}
		(a)\includegraphics[width=7.3cm]{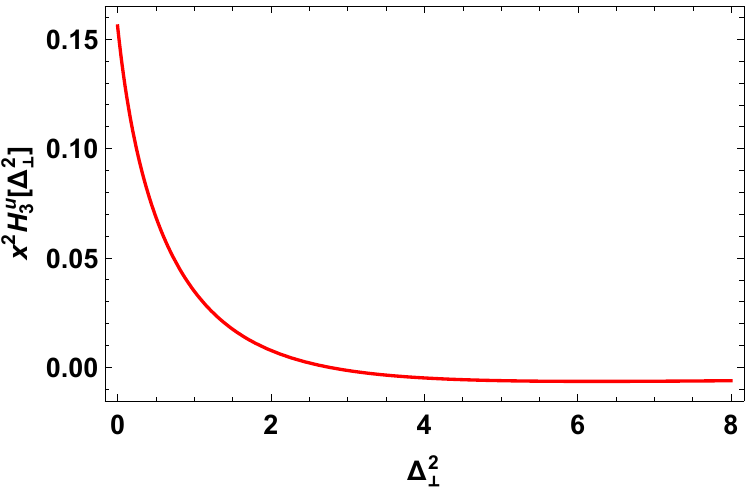}
		\hspace{0.05cm}
		(b)\includegraphics[width=7.3cm]{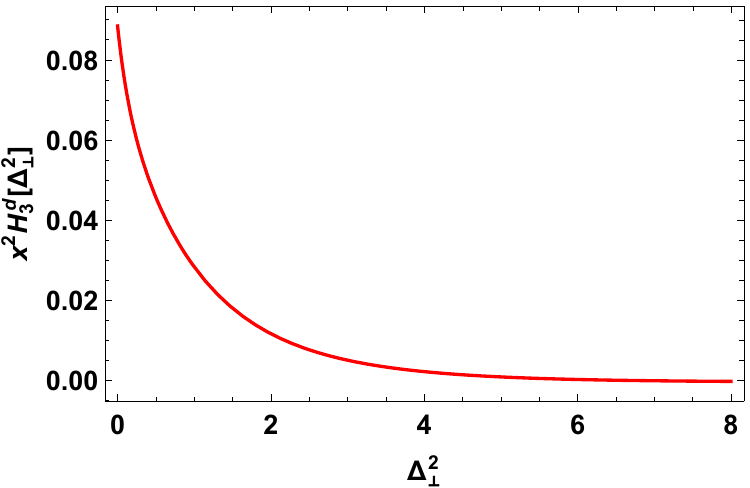}
		\hspace{0.05cm}
		(c)\includegraphics[width=7.3cm]{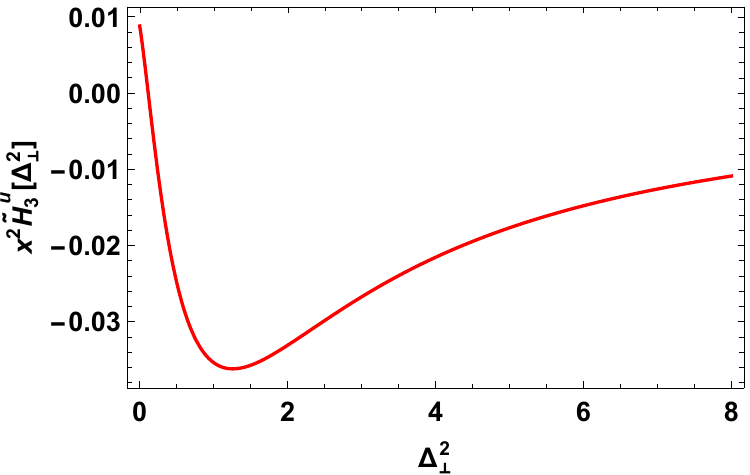}
		\hspace{0.05cm}
		(d)\includegraphics[width=7.3cm]{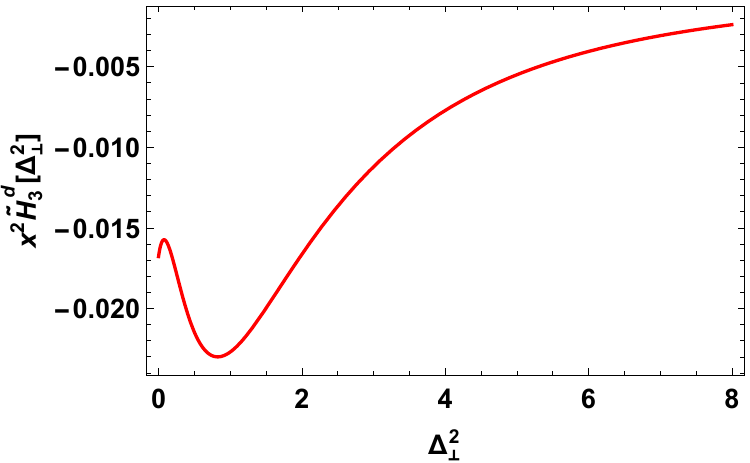}
		\hspace{0.05cm}
		(e)\includegraphics[width=7.3cm]{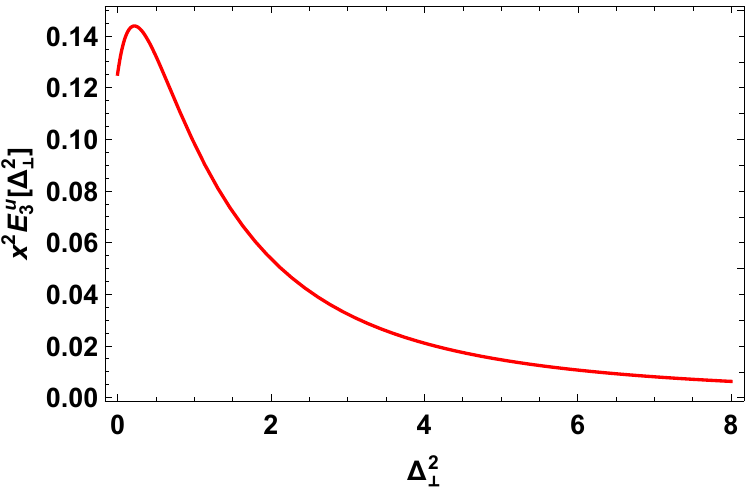}
		\hspace{0.05cm}
		(f)\includegraphics[width=7.3cm]{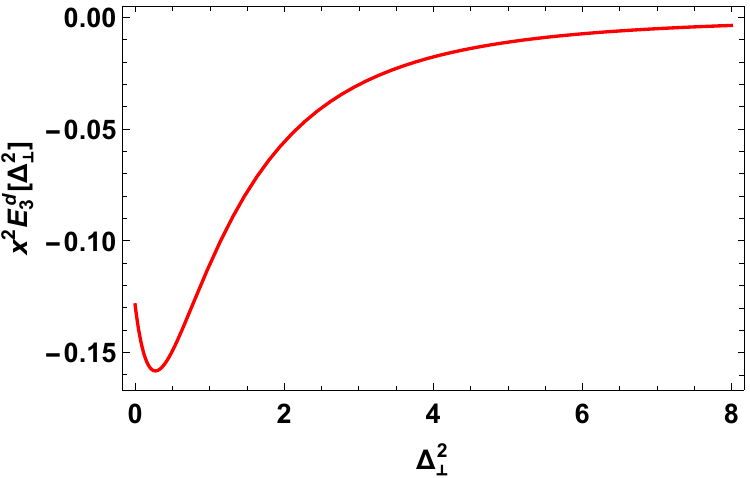}
		\hspace{0.05cm}\\
	\end{minipage}
	\caption{\label{fig2dffvd} (Color online) Twist-4 chiral-even FFs 		
		$x^2 H_{3}^{\nu}(\Dp^2)$,
		$x^2 \tilde{H}_{3}^{\nu}(\Dp^2)$ and
		$x^2 E_{3}^{\nu}(\Dp^2)$
		plotted with respect to $\Dp^2$. The left and right column correspond to $u$ and $d$ quarks sequentially.
	}
\end{figure*}

When GPDs are integrated over the longitudinal momentum fraction $x$, FFs are obtained (see Fig. \ref{figtree}). Very little work has been done on higher twist FFs, compared to twist-2 FFs \cite{Lorce:2011dv} that's why no particular symbol has been ascribed to them. To avoid any element of doubt, we will abstain from defining new symbols for them and associate them with their parent GPD symbol. To study the functioning of chiral-even FFs, we have plotted them in Fig. \ref{fig2dffvd} for $u$ and $d$ quarks in the left and right column respectively. In Fig. \ref{fig2dffvd} (a) and \ref{fig2dffvd} (b), the FF $x^2 H_{3}^{\nu}(\Dp^2)$ has been plotted for both flavors of struck quark. With the rise in the total momentum transfer $t$, the exponential fall is observed for both struck quark flavors. This suggests the exponential decreases in the possibility of obtaining the polarization corresponding to FF $x^2 H_{3}^{\nu}(\Dp^2)$ and hence negligible possibility at high momentum transfers.
\par
The FF $x^2 \tilde{H}_{3}^{\nu}(\Dp^2)$ has been plotted in Fig. \ref{fig2dffvd} (c) and \ref{fig2dffvd} (d) for $u$ and $d$ quarks respectively. For $u$ quarks, with an increase in the momentum transfer the magnitude of FF rises exponentially and reaches an 	all-time maximum at $\Dp^2=1.26$ and afterward it decreases slowly to reach the horizontal axis at a very large momentum transfer. This remnant curve indicates the high chance of obtaining this FF over other chiral-even FFs at large momentum transfer $t$. For $d$ quarks, a similar type of trend is obtained but here the FF falls to zero more rapidly. This tells at large momentum transfer $t$, the struck quark will most probably be $u$ where the polarization is most likely to be $x^2 \tilde{H}_{3}^{\nu}(\Dp^2)$ correspondent.  
\par
For the study of FF $x^2 E_{3}^{\nu}(\Dp^2)$, it has been plotted in Fig. \ref{fig2dffvd} (e) and \ref{fig2dffvd} (f) for both the flavors of struck quark. With the increase in $\Dp^2$, the amplitude of FF increases (decreases) to reach maxima (minima) and after that, it decreases (increases) exponentially to reach the horizontal axis at $\Dp^2=0.22~(0.27)$. This means that for the Dirac matrix structure $\gamma^-$, merely the change of proton polarization from unpolarized to transversely polarized leads to the need for a slightly high momentum transfer $t$, for the nearly same possibility of detecting the respective form factor.  
\subsection{Impact parameter dependent parton distributions}\label{ssipdpds}
\begin{figure*}
	\centering
	\begin{minipage}[c]{0.98\textwidth}
		(a)\includegraphics[width=7.3cm]{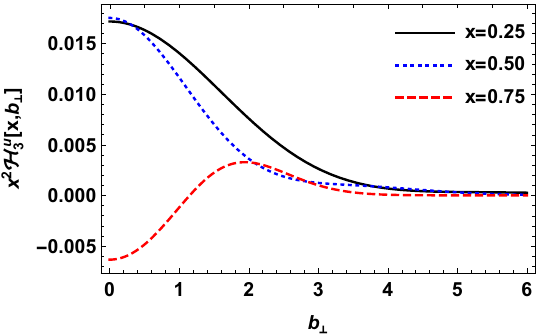}
		\hspace{0.05cm}
		(b)\includegraphics[width=7.3cm]{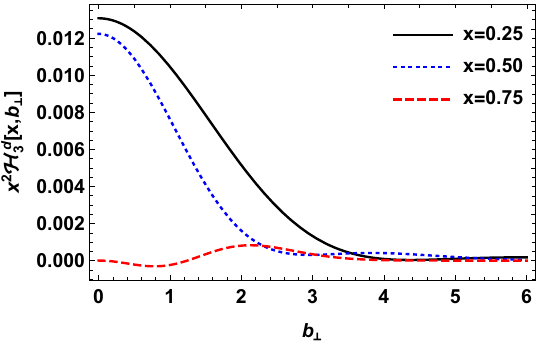}
		\hspace{0.05cm}
		(c)\includegraphics[width=7.3cm]{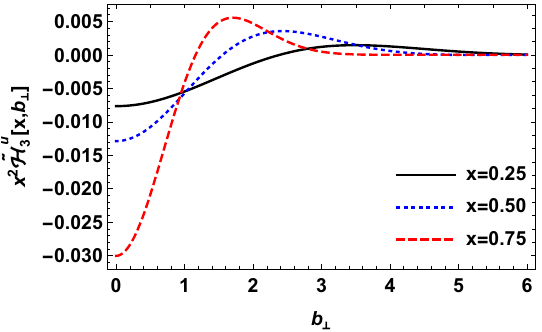}
		\hspace{0.05cm}
		(d)\includegraphics[width=7.3cm]{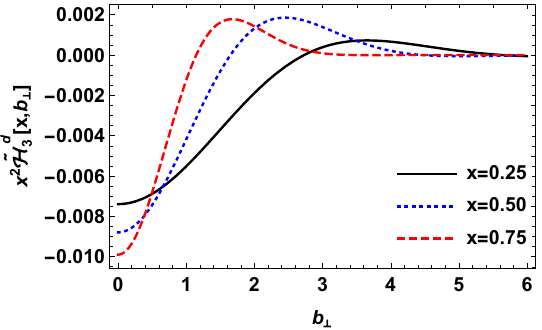}
		\hspace{0.05cm}
		(e)\includegraphics[width=7.3cm]{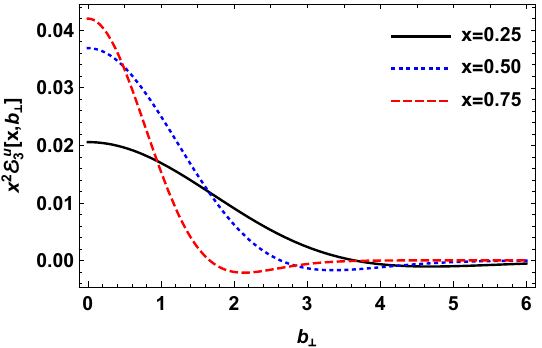}
		\hspace{0.05cm}
		(f)\includegraphics[width=7.3cm]{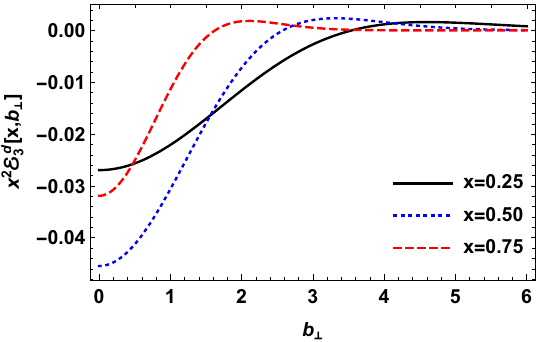}
		\hspace{0.05cm}\\
	\end{minipage}
	\caption{\label{fig2dfvb} (Color online) The Fourier transformed chiral-even twist-4 GPDs 		
		$x^2 \mathcal{H}_{3}^{\nu}(x,\bf{b_\perp})$,
		$x^2 \mathcal{\tilde{H}}_{3}^{\nu}(x,\bf{b_\perp})$ and
		$x^2 \mathcal{E}_{3}^{\nu}(x,\bf{b_\perp})$
		plotted with respect to $\bf{b_\perp}$ at various fixed values of $x$. The left and right column correspond to $u$ and $d$ quarks sequentially.
			}
\end{figure*}

The GPDs in impact parameter space, also called impact parameter dependent parton distributions (IPDPDFs), provide a different perspective on the spatial distribution of partons within hadrons \cite{Burkardt:2000za}. These distributions offer a different viewpoint on the spatial distribution of partons within hadrons by describing the probability amplitude of finding a parton at a specific transverse distance from the hadron's center. A 2-D Fourier transform in $\Dp$ is performed to obtain IPDPDFs as \cite{Burkardt:2002hr}
\begin{equation}
\mathcal{X^{\nu}}(x,\bfb)=\frac{1}{(2\pi)^2} \int d^{2}\Dp e^{-i b_{\perp}\cdot\Dp}X^{\nu}(x,\Dp^2), \\
\end{equation}
where $X^{\nu}(x,\Dp^2)$ corresponds to twist-4 GPD $H_{3}^{\nu}(x,\Dp^2)$, $\tilde{H}_{3}^{\nu}(x,\Dp^2)$ and
$E_{3}^{\nu}(x,\Dp^2)$. In Fig. \ref{fig2dfvb}, the chiral-even twist-4 GPDs in impact parameter space have been plotted with respect to the transverse distance $b_\perp$ and at certain fixed values of longitudinal momentum fraction $x$ for both possibilities of struck quark flavor. The left and right column corresponds to the struck quark being $u$ and $d$ respectively.
The plot of $x^2 \mathcal{H}_{3}^{\nu}(x,\bf{b_\perp})$ suggests that for the possibility of struck quark being at the center of momentum (COM) line is maximum when it carries longitudinal fraction $x$ upto $50\%$ from the proton longitudinal momenta as shown in Fig. \ref{fig2dfvb} (a) and \ref{fig2dfvb} (b). For this polarization, it has been observed that longitudinal momenta above this fraction are also existent in $u$ quarks but for $d$ quarks it is negligible. Also, the large fraction $x$ corresponds to the case when the distribution is more concentrated towards the COM line. $x^2 \mathcal{\tilde{H}}_{3}^{\nu}(x,\bf{b_\perp})$ plot consists of nodes prevailing at every plotted value of longitudinal momentum fraction $x$ as shown in Fig. \ref{fig2dfvb} (c) and \ref{fig2dfvb} (d). The switch in magnitude may be ascribed to the reversal in the direction of polarization and it seems to be linked proportionally to the amount of fraction $x$ smacked quark carries. Some fascinating results have been shown by the Fourier transformed transversely polarized proton GPD $x^2 \mathcal{E}_{3}^{\nu}(x,\bf{b_\perp})$ in Fig. \ref{fig2dfvb} (e) and \ref{fig2dfvb} (f). For the $u$ quarks, larger the longitudinal momentum fraction $x$, larger is the possibility of the quark being near the COM line at the plotted values of $x$. However, for $d$ quarks, the maximum concentration around the COM line is at the longitudinal fraction being half.

\section{Conclusion}\label{seccon}
In this paper, we have discussed the proton GPDs at twist-4 using the LFQDM framework. By evaluating it to the parameterization equations, we were able to decode the unintegrated quark-quark GPD correlator corresponding to the twist-4 Dirac matrix structure and, as a result, obtain the explicit equations for the chiral-even twist-4 proton GPDs. We have formulated explicit equations for GPDs on both the $u$ and $d$ struck quark scenarios, utilizing information from the scalar and vector diquark components, while considering skewness value $\xi$ to be zero. We have utilized comprehensive $2$-D and $3$-D graphs to illustrate how they rely on both longitudinal momentum fraction $x$ and momentum transfer $t$. We investigate the relevant PDFs, TMDs, and GTMDs to harmonize their findings and relationships with other DFs. In addition, the analysis of related higher twist FFs, produced from twist-4 GPDs and are crucial for comprehending the internal structure of hadrons, has been included in this paper. Finally, the results of Fourier-transformed GPDs are displayed in impact parameter space.
\par In conclusion, research in the DVCS studies at JLab, HERMES, COMPASS, MAMI and CLAS collaboration are useful to probe the internal structure of hadrons. With the advent of new experimental infrastructure, like EICs, the research of exclusive reactions involving GPDs is anticipated to grow. These resources will go through comprehension of the subtleties of the strong force and in turn offer exact information for evaluating GPD predictions and constraints. This demands theoretical computations of GPDs at all the possible proton polarizations. Our work will not only provide the necessary requisite of understanding higher twist parton dynamics but will also pave the path for future GPD calculations. In the future, it would also be fascinating to unravel the parton dynamics beyond Wandzura-Wilczek by deciphering the quark-gluon-quark correlators.
%====================================================
\section{Acknowledgement}
H.D. would like to thank the Science and Engineering Research Board, Department of Science and Technology, Government of India through the grant (Ref No.TAR/2021/000157) under TARE scheme for financial support.

%\appendix
%
%\section{}\label{App}

%___________________________________________
%Bib copied above

%___________________________________________

%
% ****** End of file apssamp.tex ******
\end{document}